\documentclass[12pt,a4paper]{article}
\usepackage{amsmath,amssymb,hyperref,bm,bbm}
\usepackage[utf8]{inputenc}
\usepackage{graphicx}
\allowdisplaybreaks[3]

\usepackage{amsthm,amscd}
\usepackage{color}
\usepackage{mathrsfs}
\usepackage{upgreek}
\usepackage[all]{xy}

\usepackage{tikz}
\usetikzlibrary{tikzmark}
\usetikzlibrary{calc}
\usepackage{pgfplots}
\usetikzlibrary{arrows,calc,decorations.markings}

\usepackage{mathtools} 
\usepackage{comment}

\numberwithin{theorem}{section}
\numberwithin{equation}{section}
\newcommand{\W}{\mathcal{W}}
\newcommand{\Z}{\mathbb{Z}}
\newcommand{\C}{\mathbb{C}}

\newcommand{\e}{\mathrm{e}}
\newcommand{\g}{\mathfrak{g}}
\newcommand{\h}{\mathfrak{h}}
\newcommand{\slf}{\mathfrak{sl}}

\newcommand{\ad}{\operatorname{ad}}

\newcommand{\Ker}{\operatorname{Ker}}

\newcommand{\dd}{\mathrm{d}}   

\newcommand{\gb}[1]{\bar{\gamma}^{#1}}

\newcommand{\bb}[1]{\bar{\beta}_{#1}}
\newcommand{\p}{\partial}
\newcommand{\pb}{\bar{\partial}}

\DeclarePairedDelimiter{\corr}{\Big\langle}{\Big\rangle} 

\newcommand{\wrt}{with respect to}

\begin{document}
\begin{titlepage}
		
\renewcommand{\thefootnote}{\fnsymbol{footnote}}
\begin{flushright}
\begin{tabular}{l}
YITP-20-22\\
\end{tabular}
\end{flushright}
		
\vfill
\begin{center}
			
			
\noindent{\large \textbf{Correspondences among CFTs with different}}
			
\medskip

\noindent{\large \textbf{W-algebra symmetry}}
			
\vspace{1.5cm}

\noindent{Thomas Creutzig$^{a}$\footnote{E-mail: creutzig@ualberta.ca},  Naoki Genra$^{a}$\footnote{E-mail: genra@ualberta.ca}, Yasuaki Hikida$^b$\footnote{E-mail: yhikida@yukawa.kyoto-u.ac.jp} and Tianshu Liu$^b$\footnote{E-mail: tianshu.liu@yukawa.kyoto-u.ac.jp}}
\bigskip

\vskip .6 truecm
\centerline{\it $^a$Department of Mathematical and Statistical Sciences, University of Alberta,} \centerline{\it Edmonton, Alberta T6G 2G1, Canada}
\medskip
\centerline{\it $^b$Center for Gravitational Physics, Yukawa Institute for Theoretical Physics,}
\centerline{\it  Kyoto University, Kyoto 606-8502, Japan}
			
\end{center}
		
\vfill
\vskip 0.5 truecm

\begin{abstract}

W-algebras are constructed via quantum Hamiltonian reduction associated with a Lie algebra $\g$ and an $\slf(2)$-embedding into $\g$.
We derive correspondences among correlation functions of theories having different W-algebras as symmetry algebras. These W-algebras are associated to the same $\g$ but distinct $\slf(2)$-embeddings. 

For this purpose, we first explore different free field realizations of W-algebras and then
generalize previous works on the path integral derivation of correspondences of correlation functions.
For $\g=\slf(3)$, there is only one non-standard (non-regular) W-algebra known as the Bershadsky-Polyakov algebra.
We examine its free field realizations and derive correlator correspondences involving the WZNW theory of $\slf(3)$, the Bershadsky-Polyakov algebra and the principal $W_3$-algebra.
There are three non-regular W-algebras associated to $\g=\slf(4)$. We show that the methods developed for $\g=\slf(3)$ can be applied straightforwardly.
We briefly comment on extensions of our techniques to general $\g$.

\end{abstract}
\vfill
\vskip 0.5 truecm
		
\setcounter{footnote}{0}
\renewcommand{\thefootnote}{\arabic{footnote}}
\end{titlepage}
	
\newpage
	
\tableofcontents

\section{Introduction}

Two-dimensional conformal field theories (CFTs) admit Virasoro symmetry, which allows to study these theories to a large extent due to this infinite dimensional symmetry. The Virasoro algebra is generated by a spin-2 current, and extended algebras can be constructed by adding higher spin currents. These algebras are called W-algebras. The standard construction of W-algebras is via Hamiltonian reduction associated with a Lie algebra $\g$ and an $\slf(2)$-embedding, see, e.g., \cite{Arakawa2017, Bouwknegt:1992wg} for reviews for mathematicians respectively physicists.
For the case of $\g=\slf(N)$, $\slf(2)$-embeddings are labeled by partitions of the integer $N$, and each partition leads to a different algebra. The partition $N=N$ corresponds to so-called principal (or regular) embedding of $\slf(2)$, which yields the W$_N$-algebra. The W$_N$-algebras has a spin-$s$ current for each $s=2,3,\ldots,N$. Furthermore, the case with partition $N=1+\cdots + 1$ corresponds to the $\slf(N)$ current algebra. Except for these two special cases, W-algebras have not been fully explored yet.

W-algebras appear in many contexts of theoretical physics.
For instance, sub-sectors of four-dimensional $SU(N)$ gauge theories are claimed to be organized by W-algebras \cite{Alday:2009aq,Wyllard:2009hg}.
In particular, 
non-regular W-algebras appear by inserting surface operators in four-dimensional gauge theories \cite{Alday:2010vg,Kozcaz:2010yp,Wyllard:2010rp,Wyllard:2010vi}. 
Moreover,  various W-algebras arise as the asymptotic symmetry of three-dimensional higher spin gravities by adopting generic gravitational sectors \cite{Henneaux:2010xg,Campoleoni:2010zq}.
Non-regular W-algebras play important roles in holographic dualities \cite{Creutzig:2018pts,David:2019bmi,Creutzig:2019qos,Creutzig:2019wfe} generalizing the original proposal of \cite{Gaberdiel:2010pz} with regular W-algebras.

In general, W-algebras are central in $S$-duality which is very closely related to the mathematics of quantum geometric Langlands duality \cite{Kapustin:2006pk}. 
The best known such duality is Feigin-Frenkel duality between the principal W-algebra of $\g$ at level $k$ and the principal W-algebra of the dual Lie algebra ${}^L\g$ at dual level ${}^Lk$ \footnote{The dual level satisfies $r^\vee (k+h^\vee)({}^Lk+{}^Lh^\vee)=1$ with $r^\vee$ the lacity of $\g$ and $h^\vee, {}^Lh^\vee$ the dual Coxeter numbers of $\g$ and ${}^L\g$.}. However $S$-duality conjectures many more dualities between non-principal W-algebras and W-superalgebras \cite{Frenkel:2018dej, Gaiotto:2017euk, Creutzig:2017uxh, Creutzig:2018ltv}. For example it has just been proven that there is a Kazama-Suzuki coset type correspondence between subregular W-algebras of type A and B and principal W-superalgebras of type $\slf(N|1)$ and $\mathfrak{osp}(2|2N)$ \cite{CGN}. While $S$-duality and the quantum geometric Langlands duality are concerned with true dualities, that is true matchings of correlation functions we are concerned with correspondences. Dualities appear if one considers correlation functions consisting of degenerate fields only. Mathematically degenerate fields should be thought of as corresponding to ordinary modules of the W-algebra and matching of correlation functions should be viewed as an equivalence of underlying tensor categories (see Conjecture 6.4 of \cite{Aganagic:2017smx} and \cite{Creutzig:2020zvv, Creutzig2018FusionCF} for  proofs of cases).  The generic field of a W-algebra is however not degenerate, but non-degenerate and we are concerned with correlation functions of fields of this type. In this case one gets correspondences, i.e. the correlation function on one side coincides with the one on the other side, but with extra degenerate field insertions.

In this paper, we derive new relations among correlation functions of theories with the symmetry of W-algebras associated with different partitions of $N$. 
Making use of the relations, we can deduce correlation functions from well-studied ones with the W$_N$-algebra or $\slf(N)$ current algebra.
The simplest example is given by a relation between $\slf(2)$ current algebra and Virasoro algebra, 
and it is called the Ribault-Teschner relation \cite{Ribault:2005wp}.
In \cite{Hikida:2007tq}, the relation was re-derived in path integral formulation, and the method allows us to derive new correspondences of correlation functions \cite{Hikida:2007sz,Creutzig:2011qm,Creutzig:2010zp,Creutzig:2015hla}.
The method of the previous works is mainly restricted to relations between $\slf(N)$ current algebra (or its superalgebra counterpart) and W-algebra corresponding to the partition $N=2 + 1 + 1 + \cdots + 1$. In order to go further and to understand relations involving other W-algebras the previous method needs to be improved and this is the aim of this paper, i.e. we extend the previous analysis by deriving new relations among more generic W-algebras. The key insight is our better understanding of different free field realizations of W-algebras.

In the previous works, we start from $\slf(N)$ Wess-Zumino-Novikov-Witten (WZNW) model with the symmetry of $\slf(N)$ current algebra. 
We use a first order formulation of the model, which corresponds to a free field realization of the current algebra.
Integrating out some of the free fields, we end up with a theory with W-algebra symmetry.
Here we would like to consider a theory with non-regular W-algebra symmetry to obtain new correspondences of correlators.
We heavily utilize free field realizations of generic W-algebras analyzed in \cite{Genra1,Genra2, CGN}.
A main point here is that there are several free field realizations of each W-algebra,
and ``nice'' correspondences can be derived by choosing convenient realizations.
We can also obtain simpler types of correspondences by putting restrictions on momenta of vertex operators inserted.

For $\g = \slf(3)$, there is only one non-regular embedding of $\slf(2)$ corresponding to the partition $3 = 2+1$.
The W-algebra labeled by the partition is known as Bershadsky-Polyakov (BP)-algebra \cite{Polyakov:1989dm,Bershadsky:1990bg}. One type of free field realization was already given in \cite{Bershadsky:1990bg}, but another type is possible by using the screening charges of \cite{Genra1}.
With the sets of screening charges, we explicitly write down the generators in terms of free fields. Making use of the expressions, we construct vertex operators transforming in representations of the BP-algebra and obtain a map among correlation functions provided by two types of  free field realizations.
We then derive correlator relations among theories with different W-algebra symmetry.
The relation between $\slf(3)$ current algebra and BP-algebra were already obtained in \cite{Creutzig:2015hla}, but here we derive the relation in a slightly different way to make our strategy clearer.
We then derive a relation between BP-algebra and W$_3$-algebra by making use of new free field realization of BP-algebra. 
We obtain more correspondences by putting restrictions on momenta of vertex operators.

We further explore examples with $\g =  \slf(4)$.
For $\slf(4)$, there are three types of non-regular W-algebras corresponding to the partitions $4=3+1$, $4 = 2 + 2$, and $4 = 2 + 1 +1 $. Even though the number of non-regular type increases, we show that the technique developed for $\slf(3)$ can be directly applied. Several types of screening charges can be constructed for each W-algebra  \cite{Genra1,Genra2}, and the explicit expressions of generators are obtained by utilizing the screening charges. We further derive new correlator correspondences by applying new free field realizations.
We also generalize the analysis to certain W-algebras associated to $\slf(N)$.

The organization of this paper is as follows.
In the next section, we first express the generators of BP-algebra in terms of two types of free field realizations and then relate the two descriptions of correlation functions.
In section  \ref{sec:corrsl3}, we derive several new correspondences among correlation functions of theories with the symmetry of W-algebras with $\slf(3)$. In particular, we make use of  new free field realization of BP-algebra.
In section \ref{sec:freesl4}, we write down the generators of three types of non-regular W-algebras with $ \slf(4)$ in terms of free fields. In section \ref{sec:corrsl4}, we obtain new correspondences among correlation functions by applying the free field realizations.
Section \ref{sec:conclusion} is devoted to conclusion and future problems.
In appendix \ref{sec:screening}, we explicitly write down screening charges for non-regular W-algebras with $\slf(4)$ and $\mathfrak{so}(5)$.
In appendix \ref{sec:corrslN}, we apply our prescription to obtain new relations among several W-algebras associated to $\slf(N)$.

\section{Free field realizations of BP-algebra}
\label{sec:FreeBP}

In this section, we examine free field realizations of BP-algebra as the simplest but non-trivial example of a non-regular W-algebra. In terms of free fields, generators are given by operators commuting with the set of the screening charges.

It is in general a difficult task to come up with a set of screening operators acting on some free field algebra such that its joint kernel is precisely the algebra of interest. In the case of the  BP-algebra and W-algebras associated to Lie algebras in general one however knows how to obtain these screening charges. 
The W-algebras are defined as homologies of complexes associated to the affine vertex algebras \cite{KRW} and it is possible to show that these homologies are isomorphic to the kernel of certain screening operators acting on some free field algebra \cite{Genra2}. The case of interest to us, that is the BP-algebra, and subregular W-algebras of  $\slf(n)$ in general has been conjectured by Feigin and Semikhatov \cite{Feigin:2004wb} and a derivation is given in  \cite[Section 3.2]{CGN}. We will give details on screening realizations of W-algebras associated to $\slf(4)$ and $\mathfrak{so}(5)$ in appendix \ref{sec:screening}. Note that the characterization of the W-algebra as a homology has the advantage that one can determine all generating fields and their conformal weights. For example the subregular W-algebra of $\slf(n)$ has $n+1$ generating fields of conformal weights $1, 2, \dots, n-1$ and $\frac{n}{2}, \frac{n}{2}$. Moreover it is enough to find the two fields of conformal weight $\frac{n}{2}$, since these two fields generate the complete W-algebra under operator products \cite{Creutzig:2020zaj}.

It turns out that there are two sets of screening operators, and hence two types of  expressions for the generators are obtained. We construct vertex operators respecting the BP-algebra for each free field realization.
If vertex operators transform in the same way under the BP-algebra, then correlation functions should be the same for the two free field realizations up to normalization of vertex operators.

\subsection{BP-algebra and its generators}

The BP-algebra can be obtained from a Hamiltonian reduction of $\slf(3)$ current algebra associated with the unique non-regular $\slf(2)$-embedding  \cite{Bershadsky:1990bg}.
The algebra is generated by a spin-one current $H(z)$, two spin-3/2 bosonic currents $G^\pm (z)$ and the energy-momentum tensor $T(z)$. The operator product expansions (OPEs) among them are given as
\begin{align}
 & T(z) T(w) \sim \frac{c/2}{(z -w)^4} + \frac{2 T(w)}{(z -w)^2} + \frac{\partial T(w) }{z-w} \, ,  \nonumber  \\
 &T(z) G^\pm (w) \sim \frac{\frac{3}{2} G^\pm (w)}{(z -w)^2} + \frac{\partial G^\pm (w)}{z-w} \, , \quad
  T(z) H(w) \sim \frac{H(w)}{(z-w)^2} + \frac{\partial H(w) }{z-w} \, ,\nonumber  \\
&H (z) H(w) \sim  - \frac{ (2 k -3)/3 }{(z -w)^2} \, , \quad
H (z) G^\pm (w)  \sim \pm  \frac{ G^\pm (w)   }{z -w} \, , \\
&G^+ (z) G^- (w) \sim \frac{(k-1)(2 k -3)}{(z-w)^3} - \frac{ 3 (k-1) H(w)}{(z-w)^2} \label{OPEs} \nonumber  \\
&\qquad \qquad \qquad+ 
\frac{ 3 (HH) (w) + (k-3) T (w) - \frac{3}{2}(k-1) \partial H(w) }{z-w} \, . \nonumber 
\end{align}
The central charge is 
\begin{align}
c = 6 (k-3)  + 25 + \frac{24 }{ k-3 } \, ,
\end{align}
where $k$ is the level of $\slf(3)$ current algebra.%
\footnote{In mathematical literature, the affine algebra level is usually set as $-k$ instead of $k$. \label{-k}}

We would like to realize the generators of BP-algebra in terms of free fields.
We introduce two free bosons $\varphi_i$ $(i=1,2)$ satisfying
\begin{align}
\varphi_i (z) \varphi_j (w) \sim - G_{ij} \log (z -w) \, ,
\end{align}
where
\begin{align}
G_{ij} = 
\begin{pmatrix}
 2 & -1 \\
-1 & 2  
\end{pmatrix} \, , \quad
G^{ij} = 
\begin{pmatrix}
 2/3 & 1/3  \\
1/3 & 2/3
\end{pmatrix} \, . \label{Gmatrices0} 
\end{align}
The indices of $\varphi_i$ can be raised and lowered  by these matrices.
We also introduce a bosonic ghost system $(\gamma,\beta)$ satisfying
\begin{align}
\gamma (z) \beta (w) \sim \frac{1}{z - w}  \, .
\end{align}
The expression of the generating fields in terms of these free fields can be found using the fact that the generating fields commute with screening charges. 

\bigskip

\noindent {\bf The first realization.}
A set of screening operators can be found in \cite{Bershadsky:1990bg} as
\begin{align}
\mathcal{S}_1 = \oint dz \mathcal{V}_1 (z)\, , \quad
\mathcal{S}_2 =\oint dz \mathcal{V}_2 (z)\label{screening1}
\end{align}
with
\begin{align}
\mathcal{V}_1 = e^{b \varphi_1}   \gamma \, , \quad
\mathcal{V}_2 =  e^{b \varphi_2} \beta   \, , \label{screening2}
\end{align}
where a new parameter $b$ is introduced as
\begin{align}
b = \frac{1}{\sqrt{k -3}} \, .
\end{align}
The generators commuting with \eqref{screening1}, \eqref{screening2} are given by (see \cite{Bershadsky:1990bg,Creutzig:2015hla})
\begin{align}
\begin{aligned}
& T = - \frac12 G^{ij} \partial \varphi _i \partial \varphi _j  + \frac{(k-1)b}{2} (\partial ^2 \varphi_1 +\partial ^2 \varphi_2 ) + \frac12 ( \gamma \partial \beta - \partial \gamma \beta) \, , \\
&H = \frac{1}{3b} ( \partial \varphi_1 - \partial \varphi_2 ) - \gamma \beta \,  , \quad
 G^+ = - \frac{1}{b} \partial \varphi_2 \gamma - \gamma \gamma  \beta + (k-1 ) \partial \gamma \, , \\
& G^- = \frac{1}{ b} \partial \varphi_1 \beta  - \gamma \beta \beta - (k -1) \partial \beta \, .
\end{aligned} \label{freeBP1}
\end{align}
Here and in the following, the normal ordering prescription is assumed for the products of free fields.
The background charges for $\varphi_i$ are set such that the conformal dimensions of screening charges are one.
With the energy-momentum tensor $T$, the conformal dimensions of $(\gamma,\beta)$ are $(1/2,1/2)$. 
We checked that the generators satisfy the OPEs in \eqref{OPEs}.

\bigskip

\noindent {\bf The second realization.}
Screening charges of free field realizations for generic W-algebras were explored in \cite{Genra1,Genra2}.
In particular, we can see that there is another set of screening operators as
\begin{align}
\mathcal{V}_1 =  e^{b \varphi_1} \, , \quad
\mathcal{V}_2 =  e^{b \varphi_2} \beta   \, . \label{screening3}
\end{align}
We find generators commuting the screening charges with \eqref{screening3} as
\begin{align}
\begin{aligned}
& T = - \frac12 G^{ij} \partial \varphi_i \partial \varphi_j + \frac{(k-1)b}{2} \partial^2 \varphi_1 + b \partial^2 \varphi_2 - \frac12  \gamma \partial \beta - \frac{3}{2} \partial \gamma \beta \, , \\
&H = \frac{1}{3b}  \partial \varphi_1  +  \frac{2}{3 b} \partial \varphi_2 + \gamma \beta \,  , \quad
 G^+  = \beta \, , \\
& G^- = - b^{-1} \partial \varphi_1 \gamma  \gamma  \beta-2 b^{-1}  \partial \varphi_2 \gamma \gamma \beta +  b^{-1} \left(k- 1\right) \partial \varphi_1 \partial \gamma + b^{-1} \left(2  k-2 \right) \partial \varphi_2 \partial \gamma \\
& \qquad + b^{-1} \left(k-2  \right) \partial ^2 \varphi_2 \gamma +(3-k) \partial \varphi_1 \partial \varphi_2 \gamma + (3-k) \partial \varphi_2 \partial \varphi_2 \gamma  +(k-3) \gamma  \gamma  \partial \beta  \\
& \qquad + (3 k-3) \partial \gamma \gamma \beta  - \gamma \gamma \gamma \beta \beta + \left(-k^2+\frac{5 k}{2}-\frac{3}{2}\right) \partial ^2 \gamma  \, . 
\end{aligned} \label{freeBP2}
\end{align}
We have checked that they satisfy \eqref{OPEs}.
Note that the conformal dimensions of $(\gamma,\beta)$ are $(-1/2,3/2)$, and it is consistent with $G^+ = \beta$.

\subsection{Vertex operators}
\label{sec:vertex}

In the previous subsection, we have written down the generators of BP-algebra in terms of free fields.
In this subsection, we introduce vertex operators and examine the action of generators to them.
Since the conformal dimensions of $G^{\pm}$ are not integer, it is convenient to redefine (or twist) the energy-momentum tensor as
\begin{align}
T(z) \to T_t (z) = T(z) + \frac12 \partial H (z) \, . \label{twist}
\end{align}
In other words, we work with the Ramond sector.
The conformal dimensions of $G^+$ and $G^-$ are one and two, respectively, with respect to the twisted energy-momentum tensor.
The generators now have integer expansion modes:
\begin{align}
\begin{aligned}
 &T_t (z) = \sum_{n\in\mathbb{Z}} \frac{L_n}{z^{n+2}}\, , \quad H (z) = \sum_{n\in\mathbb{Z}}\frac{H_n}{z^{n+1}}  \, ,\\
  &G^+(z) = \sum_{n\in\mathbb{Z}} \frac{G^+_n}{z^{n+1}} \, , \quad
G^- (z) = \sum_{n\in\mathbb{Z}} \frac{G^-_n}{z^{n+2}}.
\end{aligned}
\end{align}

With the twisted energy-momentum tensor, the conformal dimensions of $(\gamma,\beta)$ are given by $(0,1)$ for  both free field realizations.
Thus, it is natural to introduce vertex operators of the form
\begin{align}
V_{j,s}(\mu|z) = e^{\mu \gamma } e^{j \varphi_1 + s \varphi_2 } \, . \label{vosl3}
\end{align}
From the vertex operators, we define states
\begin{align}
 | j ,s ; \mu \rangle \equiv \lim_{z \to 0} V_{j,s}(\mu|z)  | 0 \rangle \, .
\end{align}
Notice that these states are primary with respect to the twisted algebra and satisfy
\begin{align}
L_n  | j ,s ; \mu \rangle =
G^\pm_n  | j ,s ; \mu \rangle  =
H_n | j ,s ; \mu \rangle & =  0
\end{align}
for $n > 0$.
Moreover, the action of zero modes can be written as
\begin{align}
\begin{aligned}
L_0  | j ,s ; \mu \rangle & = - \mathcal{D}_L  | j ,s ; \mu \rangle \, , \\
G^\pm_0  | j ,s ; \mu \rangle & = - \mathcal{D}_\pm  | j ,s ; \mu \rangle \, , \\
H_0  | j ,s ; \mu \rangle & = - \mathcal{D}_H  | j ,s ; \mu \rangle  \label{zeroactionsl3}
\end{aligned}
\end{align}
with differential operators $\mathcal{D}_L ,\mathcal{D}_\pm,\mathcal{D}_H$.
The expressions of differential operators can be read off from the action of generators to the vertex operators \eqref{vosl3}. 
One of the differential operators $\mathcal{D}_L $ is the (minus of) conformal weight as  
\begin{align}
\Delta = - \mathcal{D}_L = -j^2+ j \left( s+ b ( k-2) \right)+s \left(b -s\right) \, , \label{cw}
\end{align}
which is common for  both free field realizations.

For the first realization, we found the expression of generators as in \eqref{freeBP1}.
Acting with the generators  on the vertex operators in \eqref{vosl3}, we obtain differential operators as
\begin{align}
\begin{aligned}
\mathcal{D}_H &=-  b^{-1} (s-j)-\mu  \frac{\partial }{\partial \mu } \, , \\
\mathcal{D}_+ &=b^{-1} (j-2 s) \frac{\partial }{\partial \mu }-\mu  \frac{\partial ^2 }{\partial \mu  ^2}\, , \\
\mathcal{D}_- &= \mu ^2 \frac{\partial }{\partial \mu }-  \mu  \left(b^{-1} (2 j-s)-k+1\right)
\end{aligned}
\end{align}
along with $\mathcal{D}_L$ in \eqref{cw}.
This kind of realization of zero-mode algebra may be found in (6.16) of \cite{deBoer:1992sy} after performing Fourier transformation from $\mu$-basis to $x$-basis as
\begin{align}
V_{j,s}(x|z) =  \int d \mu e^{\mu x } V_{j,s}(\mu|z) \, . \label{mu2x}
\end{align}
These differential operators satisfy  commutation relations
\begin{align}
 [ \mathcal{D}_H , \mathcal{D}_\pm ] = \pm \mathcal{D}_\pm \, , \quad
[\mathcal{D}_+ , \mathcal{D}_- ] = - 3  \mathcal{D}_H ^2  + (2 k -3) \mathcal{D}_H + (k-3) \mathcal{D}_L \, .
\label{zerocom}
\end{align}
We also find the differential operator
\begin{align}
\mathcal{D}_3 = \mathcal{D}_H^3 + \frac{3 - 2 k}{2} \mathcal{D}_H^2 - \frac12  \{ \mathcal{D}_+ , \mathcal{D}_- \} + (3 -k) \mathcal{D}_L \mathcal{D}_H + \frac12  \mathcal{D}_H \label{3rdCasimir0}
\end{align}
which commutes with  $\mathcal{D}_L ,\mathcal{D}_\pm,\mathcal{D}_H$.  The anti-commutator is defined as $\{ A , B \} = AB + BA$.
Thus, it can be regarded as the third-order Casimir operator, whose eigenvalue can be computed from \eqref{3rdCasimir0} to be
\begin{align}
\label{3rdCasimir1}
\mathcal{D}_3  &=  \frac{1}{2 b }  \Bigl[ j^2 \left(b^{-1}-2 (k-3) s\right) \\
& \quad +j \left(-5 b^{-1} s+k \left(2 s \left(b^{-1}+s\right)-1\right)-6 s^2+2\right)-(2 k-3) s \left( b^{-1} s-1\right)\Bigr] \, . \nonumber  
\end{align}
Combined with the eigenvalue of the second-order Casimir operator in \eqref{cw}, we see that the representation is labeled by the two parameters $(j,s)$.

For the second realization, the differential operators in the zero-mode action are found to be
\begin{align}\label{z1}
\begin{aligned}
&\mathcal{D}_H =  b^{-1} s+\mu  \frac{\partial }{\partial \mu } \, , \quad
\mathcal{D}_+ = \mu \, , \\
&\mathcal{D}_- =-(j-2 s)  \left((k-3) (j+s)-  b^{-1} (k-2)\right) \frac{\partial }{\partial \mu } \\
&\qquad -\mu  \left(-3 b^{-1} s+k-3\right) \frac{\partial ^2  }{\partial \mu  ^2 }+\mu ^2 \frac{\partial ^3 }{\partial \mu ^3 }
\end{aligned}
\end{align}
in addition to $\mathcal{D}_L$ in \eqref{cw}.
This kind of realization of zero-mode algebra may be found in (2.24) of \cite{Wyllard:2010rp}  after performing the Fourier transformation  \eqref{mu2x}.
These differential operators satisfy the same commutation relations as in \eqref{zerocom}.
We can also see that the eigenvalue of the third-order Casimir \eqref{3rdCasimir0} reduces to \eqref{3rdCasimir1} upon the substitution of \eqref{z1}.

\subsection{Map of correlation functions}
\label{sec:maps}

The two types of free field realizations for the BP-algebra described above allow correlation functions to be presented in two different forms.
For later analysis,  both types of descriptions are used, so we need a map from one description to the other.
As shown in \eqref{zeroactionsl3}, the action of generators on primary states is given by differential operators. However, the expressions for $\mathcal{D}_\pm ,\mathcal{D}_H$ are different for the two realizations. 
We have already observed that the eigenvalues of the second- and third-order Casimir operators are the same for both realizations, which means that the primary states with the same $(j,s)$ -values belong to the same representation.
In this subsection, we find new bases for vertex operators such that generators act identically for both descriptions.

It is easier to see the correspondence between the two realisations with a suitable choice of basis.
We shall move from $\mu$-basis to $m$-basis by performing Mellin transformation as 
\begin{align}
| j , s ; m \rangle ^{(1)} = \int d \mu \mu^{- m} | j , s ; \mu \rangle \, . \label{mbasis0}
\end{align}
For the first realization, the action of generators \eqref{freeBP1}, in terms of the new basis, can be read off as
\begin{align}
\begin{aligned}
&\mathcal{M}_H  ^{(1)}= 1 - m - b^{-1} ( - j + s)  \, ,  \\
&\mathcal{M}_+ ^{(1)}= - m(m - 1  - b^{-1} (j - 2 s) ) \, , \\
&\mathcal{M}_- ^{(1)}=    m - 3 + k - b^{-1} (2 j -s)  \, .
\end{aligned}\label{1stM0}
\end{align}
We define $\mathcal{M}_H^{(\alpha)},\mathcal{M}_\pm^{(\alpha)}$ by
\begin{align}
G^\pm_0  | j ,s ; m \rangle^{(\alpha)} = - \mathcal{M}_\pm ^{(\alpha)} | j ,s ; m \pm 1  \rangle ^{(\alpha)}\, , \quad
H_0  | j ,s ; m\rangle  ^{(\alpha)}= - \mathcal{M}_H ^{(\alpha)} | j ,s ; m\rangle ^{(\alpha)} \, .
\end{align}
We further change the basis as
\begin{align}
| j , s ; m \rangle ^{(2)} = (-1)^m \Gamma (m) \Gamma(m - 1  - b^{-1} (j - 2 s)) \int d \mu \mu^{-m} | j , s ; \mu \rangle \, , \label{mbasis1}
\end{align}
which leads to
\begin{align}
\begin{aligned}
&\mathcal{M}_H  ^{(2)}= 1 - m - b^{-1} ( - j + s)  \, , \quad
\mathcal{M}_+ ^{(2)}  = 1 \, , \\
&\mathcal{M}_-  ^{(2)}= 
  - (  m - 3 + k - b^{-1} (2 j -s)  ) (m -1) ( m -2  - b^{-1} (j - 2 s))  \, . \\
\end{aligned}\label{1stM}
\end{align}
In terms of vertex operators, the change of basis is defined as
\begin{align}
\Phi_{j,s;m}^{(2)} (z) =  (-1)^m \Gamma (m) \Gamma(m - 1  - b^{-1} (j - 2 s)) \int d \mu \mu^{-m}  V_{j,s} (\mu | z) \, . \label{mbasis2}
\end{align}

For the second realization with the generators \eqref{freeBP2}, 
we may change the basis as
\begin{align}
| j  , s  ; m'  \rangle^{(3)} = \int d \mu \mu^{ m ' } | j  , s  ; \mu \rangle \, .
\end{align}
From this definition, we find
\begin{align}
\begin{aligned}
&\mathcal{M}_H  ^{(3)} =  b^{-1} s - m ' - 1  \, , \quad
\mathcal{M}_+  ^{(3)} = 1  \, , \\
&\mathcal{M}_-  ^{(3)}= \left( (2 - k) b^{-1} + (k-3) (j + s)    \right) (j  - 2 s ) m '  \\
& \qquad  + \left(3 b^{-1} s -k+3\right) m ' (m ' + 1) - m ' (m ' + 1) (m ' + 2) \, . \\
\end{aligned} \label{2ndM}
\end{align}
Comparing \eqref{1stM} with \eqref{2ndM}, the coefficients $\mathcal{M}_H^{(\alpha)},\mathcal{M}_\pm^{(\alpha)}$ from the two realizations become identical if we set 
$
m ' =  m-2- b^{-1} (j-2 s)
$.
In other words, changing the basis as
\begin{align}
| j  , s  ; m  \rangle ^{(4)} = \int d \mu \mu^{  m-2- b^{-1} (j-2 s)} | j  , s  ; \mu \rangle \, , 
\end{align}
we can realize
\begin{align}
\mathcal{M}_H^{(2)} = \mathcal{M}_H^{(4)} \, , \quad \mathcal{M}_\pm^{(2)} = \mathcal{M}_\pm^{(4)} \, .
\end{align}
The corresponding vertex operators may be introduced as 
 \begin{align}
\Phi_{j,s;m}^{(4)} (z)=   \int d \mu \mu^{  m-2- b^{-1} (j-2 s)} V_{j,s} (\mu | z) \, .  \label{mbasis3}
\end{align}
Since the actions of generators on the vertex operators \eqref{mbasis2} and \eqref{mbasis3} are the same now,
correlation functions computed with vertex operators $\Phi_{j,s;m} ^{(2)}(z)$ and those with $\Phi_{j,s;m} ^{(4)}(z)$  should be the same once their normalization is properly set.

\section{Correlator relations for W-algebras from $\slf(3)$}
\label{sec:corrsl3}

In this section, we derive correspondences among correlation functions of theories with the symmetry of W-algebras 
associated with $\slf(3)$.
In subsection \ref{sec:previous}, we reduce the $\slf(3)$ WZNW model to a theory with BP-algebra symmetry in a way slightly different from \cite{Creutzig:2015hla}.
In subsection \ref{sec:BPW3}, we obtain correspondences between theories with the symmetry of BP-algebra and W$_3$-algebra using the free field realizations of BP-algebra analyzed in the previous section.
Correlators of vertex operators with restricted momenta are examined in subsection \ref{sec:restricted}. These correlators take simpler forms compared to those we studied previously.
In subsection \ref{sec:Direct} we propose several ways to obtain direct relations between $\slf(3)$ current algebra and W$_3$-algebra.

\subsection{Reduction from affine $\slf(3)$ to BP-algebra}
\label{sec:previous}

Starting from correlation functions of  the $\slf(3)$ WZNW model, we formulate the action in this first order as we did in \cite{Creutzig:2015hla}:
\begin{align}
\begin{aligned}
 S &= \frac{1}{2 \pi} \int d ^2 z \left[ \frac{G_{ij}}{2} \partial \phi^i \bar \partial \phi^j + \frac{b}{4}\sqrt{g} \mathcal{R} (\phi_1 + \phi_2) + \sum_{\alpha=1}^3 \left( \beta_\alpha \bar \partial \gamma^ \alpha + \bar \beta_\alpha \partial \bar \gamma^\alpha  \right) \right] \\
 &\quad - \frac{1}{2 \pi k} \int d^2 z \left[ e^{b \phi_1} (\beta_1 - \gamma^2 \beta_3 ) (\bar \beta_1 - \bar \gamma^2 \bar \beta_3) + e^{b \phi_2} \beta_2 \bar \beta_2 \right] \, ,
\end{aligned} \label{actionsl3}
\end{align}
where the matrix $G_{ij}$ was defined in \eqref{Gmatrices0}. Moreover, $g_{\sigma\rho} $ is the world-sheet metric, $g = \det g_{\sigma\rho}$, and $\mathcal{R}$ represents the scalar curvature.
In the path integral formulation, correlation functions can be written as
\begin{align}
 \left \langle \prod^N_{\nu = 1} V_\nu (z_\nu) \right \rangle =  \int \mathcal{D} \Phi e^{- S}  \prod^N_{\nu = 1} V_\nu (z_\nu) \, , \label{corr}
\end{align} 
where the path integral measure is
\begin{align}
 \mathcal{D} \Phi = \mathcal{D} \phi_1 \mathcal{D} \phi_2 \prod_{\alpha =1}^3 \mathcal{D}^2 \beta_\alpha \mathcal{D}^2 \gamma^\alpha \, . \label{measuresl3}
\end{align}
The vertex operators are defined as
\begin{align}
 V_\nu (z_\nu) = |\mu_1^\nu|^{4 (j_1^\nu +1)}|\mu_3^\nu/\mu_1^\nu|^{4 (j^\nu_2 +1)} e^{\mu_\alpha^\nu\gamma^\alpha-\bar{\mu}_\alpha^\nu\bar{\gamma}^\alpha} e^{2b(j^\nu_1+1)\phi_1 + 2b(j^\nu_2+1)\phi_2} \, . \label{vertexsl3}
\end{align}
The prefactors are chosen such that the final expression becomes simpler.

We would like to reduce the theory to that with BP-algebra symmetry.
In the previous section, we have seen that BP-algebra can be realized by  two free bosons $\varphi_i$ $(i=1,2)$ and a ghost system $(\gamma , \beta)$ along with proper screening charges.
In order to realize the system in terms of action, we need to add anti-holomorphic counterparts  $\bar \varphi_i$ and  $(\bar \gamma , \bar \beta)$. The free bosons in the two sectors can be formulated as a single non-chiral field as
\begin{align}
\phi_i  (z , \bar z )= \varphi_i (z) + \bar \varphi_i (\bar z ) \, . \label{dec}
\end{align}
We shall obtain the theory with BP-symmetry from the $\slf(3)$ WZNW model in the first order formulation by integrating out two sets of ghost system.

Following  \cite{Hikida:2007tq,Creutzig:2015hla}, we first integrate with respect to $\gamma^1 , \gamma^3$ and  $\bar \gamma^1 , \bar \gamma^3$, which appear only linearly in the exponent of path integral expression \eqref{corr}.
The integration over zero-modes of these fields leads to delta functions
\begin{align}
\delta ^{(2)} \left( \sum_{\nu=1}^N \mu_1^\nu \right) 
\delta ^{(2)} \left( \sum_{\nu=1}^N \mu_3^\nu\right) \, .
\end{align}
Moreover, non-zero modes provide delta functionals for $\beta_1, \beta_3$ and $\bar \beta_1 , \bar \beta_3$.
Integration with respect to these fields leads to the replacement of them by functions:
\begin{align}
&\beta_\alpha (z) = - \sum_{\nu=1}^N \frac{\mu_\alpha^\nu}{z - z_\nu} = - u_\alpha \frac{\prod_{n=1}^{N-2} (z - y_\alpha^n)}{\prod_{\nu=1}^N (z - z_\nu)} \equiv  - u_\alpha \mathcal{B}_\alpha (z , z_\nu , y_\alpha^n ) \, , \\
&\bar \beta_\alpha (\bar z) =  \sum_{\nu=1}^N \frac{\bar \mu_\alpha^\nu}{\bar z - \bar z_\nu}=  \bar u_\alpha \frac{\prod_{n=1}^{N-2} (\bar z - \bar y_\alpha^n)}{\prod_{\nu=1}^N (\bar z - \bar z_\nu)} \equiv  \bar u_\alpha  \bar{\mathcal{B}}_\alpha (\bar z ,\bar z_\nu , \bar y_\alpha^n ) 
\end{align}
with $\alpha =1,3$. 
Notice that a holomorphic 1-form possesses exactly two more poles than its zeros. Therefore, there are $(N-2)$ zeros for each $\mathcal{B}_\alpha$ or $\bar{\mathcal{B}}_\alpha$, and these zeros are positioned at coordinates $y_\alpha^n$.

The interaction terms in the action now become
\begin{align}
\begin{aligned}
\frac{1}{2 \pi k} \int d^2 z \left[ e^{b \phi_1} (u _1 \mathcal{B}_1 - \gamma^2 u_3 \mathcal{B}_3 ) (\bar u_1 \bar{\mathcal{B}}_1 - \bar \gamma^2 \bar u_3 \bar{\mathcal{B}}_3) - e^{b \phi_2} \beta_2 \bar \beta_2 \right] \, .
 \end{aligned} \label{actionsl3p}
\end{align}
In order to relate the action to one of the free field realizations of BP-algebra, we need to remove the function dependence in the interaction terms by redefining fields $\phi_1,\phi_2,\gamma^2$ and $\beta_2$.
There are several ways to do so, but here we choose to make the shifts of $\phi_1$ and $\phi_2$ as
\begin{align}
 \phi_1 +  \frac{1}{b} \log |u_1 \mathcal{B}_1| ^2  \to \phi_1 \, , \quad  \phi_2 +  \frac{1}{b} \log |u_1^{-1} u_3 \mathcal{B}_1^{-1 } \mathcal{B}_3 | ^2  \to \phi_2
\end{align}
and change $\gamma^2$ and $\beta_2$ as
\begin{align}\label{z10}
 \gamma^2 u_1^{-1}  u_3 \mathcal{B}_1^{-1}  \mathcal{B}_3 \to \gamma^2 \, , \quad
 \beta_2 u_1  u_3^{-1} \mathcal{B}_1  \mathcal{B}_3^{-1} \to \beta_2 \, . 
\end{align}
The conjugate fields $\bar \gamma^2$ and $\bar \beta_2$ are changed in the same way as in \eqref{z10}. 
The field redefinitions give rise to extra factors in the kinetic terms.
A detailed computation can be found in \cite{Hikida:2007tq,Creutzig:2015hla}.
Part of the contributions from  the kinetic terms of $\phi_1$, $\phi_2$ can be regarded as the shifts of momenta for vertex operators and the insertions of extra fields at $y_1^n,y_3^n$. The change of fields also results in the shifts of background charges and a prefactor in front of correlation function. To see the change in the correlation function due to \eqref{z10}, it is convenient to rewrite $\gamma^2$ and $\beta_2$ 
\begin{align}
\gamma ^2 (z) \simeq e^{X (z)} \eta (z)  \, , \quad 
\beta _2 (z) \simeq e^{- X (z)} \partial \xi (z) \, ,
\end{align}
where the bosonic field $X$ and the fermionic fields $\eta , \xi $ satisfy the OPEs
\begin{align}
 X (z) X(w) \sim - \log (z - w) \, , \quad \eta (z) \xi (w) \sim \frac{1}{z -w} \, .
\end{align}
With this terminology, the change of ghost system can be realized by a shift of $X$ 
\begin{equation}
X-\ln{(u_1 u_3^{-1}\mathcal{B}_1\mathcal{B}_3^{-1})}\rightarrow X.
\end{equation}

Overall, we obtain the relation among correlation functions as
\begin{align}
 \left \langle \prod^N_{\nu = 1} V_\nu (z_\nu) \right \rangle 
= |\Theta_N|^2 \delta ^{(2)} \left( \sum_{\nu=1}^N \mu_1^\nu \right) 
\delta ^{(2)} \left( \sum_{\nu=1}^N \mu_3^\nu\right) 
 \left \langle \prod^N_{\nu = 1} \tilde V_\nu (z_\nu) \prod_{n=1}^{N-2} \tilde V^{(1)} (y_1^n) \tilde  V^{(3)} (y_3^n) \right \rangle \, .
\label{sl3toBP}
\end{align} 
The new vertex operators are
\begin{align}
\begin{aligned}
&\tilde V_\nu (z_\nu)  = e^{ \mu_2 '{}^{ \nu} \gamma^2 -   \bar{\mu}  ' _2 {}^  \nu  \bar{\gamma}^2} e^{2 b (j_1^\nu + 1) \phi_1 + 2 b (j_2^\nu +1) \phi_2 + \phi^1 /b} \, , \\
& \tilde V^{(1)} (y_1^n) = e^{- \phi^1/b + \phi^2 /b} e^{X+\bar{X}} \, , \quad
 \tilde V^{(3)} (y_3^n) = e^{-  \phi^2 /b} e^{-X-\bar{X}}
\end{aligned} \label{tvertexsl3}
\end{align}
with
\begin{align}
\mu_2 '{}^{ \nu} =  \frac{u_1 \mu_1^\nu  \mu_2^\nu}{ u_3 \mu_3^\nu } \, , \qquad
\bar{\mu}_2 '{}^{ \nu} =  \frac{\bar{u}_1 \bar{\mu}_1^\nu  \bar{\mu}_2^\nu}{ \bar{u}_3 \bar{\mu}_3^\nu } \, .
\end{align}
The prefactor is
\begin{align}
\begin{aligned}
\Theta_N &= u_1 u_3 \prod_{\mu < \nu} (z_\mu - z_\nu)^{2/(3 b^2)} 
\prod_{n,\nu}((y_1^n - z_\nu) (y_3^n - z_\nu))^{-1/(3 b^2)} \\
& \quad \times
\prod_{m < n} ((y_1^m - y_1^n ) (y_3^m - y_3^n))^{2/(3 b^2)+1} \prod_{m,n} (y_3^m - y_1^n)^{-1/(3 b^2)-1} \, .
\end{aligned}
\end{align}
The left-hand side of \eqref{sl3toBP} is computed with the action 
\begin{align}
\begin{aligned}
S  &= \frac{1}{2 \pi} \int d ^2 z \left[ \frac{G_{ij}}{2} \partial \phi^i \bar \partial \phi^j + \frac{1}{4}\sqrt{g} \mathcal{R} ((b + b^{-1})\phi^1 + b \phi^2) +  \beta_2 \bar \partial \gamma^ 2 + \bar \beta_2 \partial \bar \gamma^2  \right] \\
&\quad + \frac{1}{2 \pi k} \int d^2 z \left[ e^{b \phi_1} (1 - \gamma^2 ) (1 - \bar \gamma^2 ) -  e^{b \phi_2} \beta_2 \bar \beta_2 \right] \, .
\end{aligned} \label{actionBP0}
\end{align}
We can see that the action describes the first realization with the set of screening charges \eqref{screening2} after the shift $\gamma^2 - 1  \to \gamma^2$. The conformal dimensions of $(\gamma^2,\beta_2)$  are $(0,1)$, which means that the energy-momentum tensor is twisted according to \eqref{twist}.%
\footnote{In \cite{Creutzig:2015hla}, the change of fields is made such that $(\gamma,\beta)$ left have conformal dimensions $(1/2,1/2)$. In other words, the energy-momentum tensor for the reduced theory is not twisted there.}

\subsection{Reduction from BP-algebra to W$_3$-algebra}
\label{sec:BPW3}

In the previous subsection, we examined the specific example of reducing the $\mathfrak{sl}(3)$ correlation function to that of the BP-algebra. The strategy for obtaining new correlator relations may be summarized in general as follows.
We start from an action with free kinetic terms plus interactions.
We then integrate out several sets of ghost system and perform field redefinitions to eliminate explicit coordinate dependence in the action. Carefully treating contributions from kinetic terms, we obtain a correspondence among correlators of different theories.

In this subsection, we reduce correlators of the theory with BP-algebra symmetry to $\slf(3)$ Toda field theory with W$_3$-algebra symmetry by applying the generic method.
As seen above, there are two types of free field realizations for BP-algebra, which  means that there are two theories we could start from. With the procedure described above,  one of the theories is suitable for the reduction to the W$_3$-correlator. In subsection \ref{sec:previous}, we ended up with the action corresponding to the first realization with the screening operators \eqref{screening2}.
However, with this form of the action, it is not easy to integrate $(\gamma^2,\beta_2)$ out, since $\gamma^2$ appears in an interaction term.
Fortunately, we have the second realization with the screening operators \eqref{screening3}, and 
the corresponding action can be written as%
\footnote{The prefactors in front of interaction terms can be chosen arbitrary by constant shifts of $\phi_1,\phi_2$.}
\begin{align}
\begin{aligned}
S   &= \frac{1}{2 \pi} \int d ^2 z \left[ \frac{G_{ij}}{2} \partial \phi^i \bar \partial \phi^j + \frac{1}{4}\sqrt{g} \mathcal{R} ((b + b^{-1})\phi^1 + b \phi^2) +  \beta \bar \partial \gamma + \bar \beta \partial \bar \gamma   \right] \\
&\quad +  \frac{1}{2 \pi k} \int d^2 z \left[ e^{b \phi_1} - e^{b \phi_2} \beta  \bar \beta   \right] \, . \label{actionBP2}
\end{aligned}
\end{align}
It is then possible to integrate out $(\gamma ,\beta )$ from the correlation function associated to this action, and the rest can be proceeded by applying the generic method.

We take the correlation function of the form \eqref{corr} with the action \eqref{actionBP2} and the path integral measure
\begin{align}
 \mathcal{D} \Phi = \mathcal{D} \phi_1 \mathcal{D} \phi_2 \mathcal{D}^2 \beta \mathcal{D}^2 \gamma  \, .
\end{align}
The vertex operators are
\begin{align}
 V_\nu (z_\nu) = |\mu_2^\nu|^{4 (j_2^\nu +1)} e^{\mu^\nu \gamma  -\bar{\mu}^\nu  \bar{\gamma}  } e^{2b(j^\nu_1+1)\phi_1 + 2b(j^\nu_2+1)\phi_2} \, .
\end{align}
Again, the prefactor is chosen such that the final expression becomes simpler.

The integration over $(\gamma , \beta )$
leads to $\delta^{(2)} (\sum_\nu \mu ^\nu)$ and the replacement of $\beta $ by a function as
\begin{align}
\beta (z) = - \sum_{\nu=1}^N \frac{\mu^\nu}{z - z_\nu} = - u \frac{\prod_{n=1}^{N-2} (z - y^n)}{\prod_{\nu=1}^N (z - z_\nu)} \equiv  -   \mathcal{B} (z, z_\nu , y^n )\, .
\end{align}
There is $|\mathcal{B} |^2$ in an interaction term now, and we remove it by shifting $\phi_2$ as
\begin{align}
\phi_2 +  \frac{1}{b} \log |u  \mathcal{B}  | ^2  \to \phi_2 \, .
\end{align}
Similarly to the previous case, the change of $\phi_2$ in the kinetic term contributes extra factors to the correlation function.

In the end, we arrive at the relation 
\begin{align} \label{BPtoW3}
 \left \langle \prod^N_{\nu = 1} V_\nu (z_\nu) \right \rangle 
= |\Theta_N|^2 \delta ^{(2)} \left( \sum_{\nu=1}^N \mu^\nu \right) 
 \left \langle \prod^N_{\nu = 1} \tilde V_\nu (z_\nu) \prod_{n=1}^{N-2} \tilde V (y^n) \right \rangle \, .
\end{align} 
The right-hand side is evaluated with the action of $\slf(3)$ Toda field theory 
\begin{align}
S = \frac{1}{2 \pi} \int d ^2 z \left[ \frac{G_{ij}}{2} \partial \phi^i \bar \partial \phi^j + \frac{Q_\phi}{4}\sqrt{g} \mathcal{R} (\phi_1 +  \phi_2)  + \frac{1}{k}( e^{b \phi_1} + e^{b \phi_2} ) \right]
\end{align}
with $Q_\phi = b + b^{-1}$. The vertex operators are
\begin{align}
\begin{aligned}
&\tilde V_\nu (z_\nu)  = e^{2 b (j_1^\nu + 1) \phi_1 + 2 b (j_2^\nu +1) \phi_2 + \phi^2 /b} \, ,
& \tilde V (y^n) = e^{- \phi^2 /b} 
\end{aligned}
\end{align}
and the prefactor is
\begin{align}
\begin{aligned}
\Theta_N = u^{2k/3} \prod_{\mu < \nu} (z_\mu - z_\nu)^{2/(3 b^2)} 
\prod_{n,\nu}(y^n - z_\nu) ^{-2/(3 b^2)}
\prod_{m < n} (y^m - y^n  ) ^{2/(3 b^2)}  \, .
\end{aligned}
\end{align}

\subsection{Putting restrictions on momenta}
\label{sec:restricted}

In subsection \ref{sec:previous}, we have considered the action \eqref{actionsl3} associated to the $\mathfrak{sl}(3)$ WZNW model.
The action includes $\gamma^2$ in an interaction term, which makes it difficult to integrate $\gamma^2$ out.
In this subsection, we propose a way to resolve the issue by putting restrictions on momenta of vertex operators.
As before, we start from the $\mathfrak{sl} (3)$ correlator \eqref{corr} with the action \eqref{actionsl3} and the path integral measure \eqref{measuresl3}.  
However, we use a slightly different vertex operator
\begin{align}
 V_\nu (z_\nu) = |\mu_1^\nu|^{4 (j_1^\nu +1)} e^{\mu_1^\nu\gamma^1 + \mu_2^\nu\gamma^2 - \bar \mu_1^\nu \bar \gamma^1 - \bar  \mu_2^\nu \bar \gamma^2} e^{2b(j^\nu_1+1)\phi_1 + 2b(j^\nu_2+1)\phi_2} \, .
\end{align}
Namely, we set
\begin{align}
\mu^\nu_3 = \bar{\mu}^\nu_3 = 0 \label{restriction}
\end{align}
in \eqref{vertexsl3} for all $\nu$.
In this case, integration over $\gamma^3 $ leads to delta functional
\begin{align}
\delta^{(2)} ( \bar \partial \beta_3 (z)) \, .
\end{align}
With appropriate boundary conditions, we can set
\begin{align}
\beta_3 (z) = \bar \beta_3 (\bar z) = 0 \, .
\end{align}
Substituting this into the action \eqref{actionsl3},  $\gamma^2$-dependent term disappears.

Integrating the correlator with respect to $\gamma^1$ gives
\begin{equation}\label{a04}
\beta_1(z)= - \sum_{\nu=1}^N \frac{\mu^\nu_1}{z - z_\nu}  = - u _1 \frac{\prod_{n=1}^{N-2}(z-y^n_1)}{\prod_{\nu=1}^N(z-z_\nu)} = - u _1 \mathcal{B} _1 \, , \quad
\sum_{\nu=1}^{N}\mu_1^{\nu}=0 \, .
\end{equation}
We make a shift in $\phi_1(z)$ by letting 
\begin{equation}
\phi_1 + \frac 1 b \ln| u _1 \mathcal{B} _1 |^2 \to \phi_1 ,
\end{equation}
which yields the relation
\begin{align} \label{sl3toBP2}
 \left \langle \prod^N_{\nu = 1} V_\nu (z_\nu) \right \rangle 
= |\Theta_N|^2 \delta ^{(2)} \left( \sum_{\nu=1}^N \mu_1^\nu \right) 
 \left \langle \prod^N_{\nu = 1} \tilde V_\nu (z_\nu) \prod_{n=1}^{N-2} \tilde V (y^n_1) \right \rangle \, .
\end{align} 
The right-hand side is evaluated by the action \eqref{actionBP2} with $(\gamma, \beta)$ replaced by $(\gamma^2 , \beta_2)$. The vertex operators are
\begin{align}
\begin{aligned}
&\tilde V_\nu (z_\nu)  = e^{\mu_2^\nu \gamma^2 - \bar \mu_2^\nu \bar \gamma^2 }e^{2 b (j_1^\nu + 1) \phi_1 + 2 b (j_2^\nu +1) \phi_2 + \phi^1 /b} \, ,
& \tilde V (y^n) = e^{- \phi^1 /b} 
\end{aligned}
\end{align}
and the prefactor is
\begin{align}
\begin{aligned}
\Theta_N = u^{2}_1 \prod_{\mu < \nu} (z_\mu - z_\nu)^{2/(3 b^2)} 
\prod_{n,\nu}(y^n_1 - z_\nu) ^{-2/(3 b^2)}
\prod_{m < n} (y^m_1 - y^n_1 ) ^{2/(3 b^2)}  \, .
\end{aligned}
\end{align}
As was done in the previous subsection, we can reduce the theory to $\slf(3)$ Toda field theory by further integrating $(\gamma ^2 , \beta _ 2 )$ out.

In this way, we found a version of correlator relation by putting restrictions on momenta for vertex operators of $\slf(3)$ WZNW model  as in \eqref{restriction}.
In general it is possible to obtain more new relations with similar restrictions.
In the following, we would like to clarify the features of correlator relations obtained in this manner by working on the simplest setting with $\slf(2)$.

We begin with the $\slf(2)$ WZNW model with the action
\begin{align}
S = \frac{1}{2 \pi} \int d^2 w \left[\bar \partial \phi \partial \phi + \beta \bar \partial \gamma + \bar \beta \partial \bar \gamma + \frac{b}{4} \sqrt{g} \mathcal{R} \phi - \frac{1}{k} \beta \bar \beta e^{2 b \phi}\right] \, ,
\end{align}
where we set  $b =1/\sqrt{ k-2 }$.
We consider  the vertex operators in the form
\begin{align}
V_j (\mu_\nu | z) = |\mu|^{2 j + 2} e^{\mu \gamma- \bar \mu \bar \gamma }  e^{2 b (j +1) \phi } \, .
\end{align}
We may change the basis by Mellin transformation as
\begin{align}
\Phi^j_{m , \bar m} (z)= \int \frac{d^2 \mu}{|\mu|^2} \mu^{-m} \mu^{- \bar m} V_j (\mu | z) \propto \gamma(z)^{- j - 1 +m} \bar \gamma(\bar z)^{- j - 1 + \bar m} e^{2 b (j +1) \phi (z,\bar z)} \, . \label{hwop}
\end{align}
We can see that $V_j (\mu |z)$ with $\mu = \bar \mu = 0$ can be identified with $\Phi^j_{j+1,j+1} (z)$
up to an overall normalization. The operator $\Phi^j_{j+1,j+1} (z)$ is special in the sense that it corresponds to a  highest-weight state.

Let us  consider an $N$-point function and put restrictions on momenta as
\begin{align}
\mu^\nu = \bar \mu^\nu =0  \quad (\nu = M+1 ,M+2 ,\ldots ,N) \, .
\end{align}
After integrating over  $\gamma$, we obtain the relation
\begin{align}
\beta (z) = - \sum_{\nu=1}^M \frac{\mu^\nu}{w - z_\nu}  = - u \frac{\prod_{n=1}^{M-2} (z - y^n)}{\prod_{\nu = 1}^M (z - z_\nu)} \, , \quad \sum_{\nu =1}^{M}  \mu^\nu = 0 \, .
\end{align}
Since the number of poles is now $M$, the holomorphic 1-form possesses only $M-2$ zeros located at $z = y^n$.
This implies that the number of extra insertions is $M-2$, and therefore, we can obtain a relation among an $N$-point function of $\slf(2)$ WZNW model and an $(N+M-2)$-point function of Liouville field theory.
For $N=4,M=2$, this relation is essentially the one presented in section 3.2 of \cite{Chang:2014jta} with a certain limit of parameters.
Furthermore, this relation was utilized to examine BPS conformal blocks of $\mathcal{N}=2$ superconformal field theory in  \cite{Lin:2015wcg,Lin:2016gcl}.

Similar arguments can be applied to more generic examples of $\slf(L)$ with $L=3,4,\ldots$.
An operator with $\mu_\alpha= 0$ for some $\alpha$ generically corresponds to a highest-weight state,
and correlator relations obtained as in this subsection should be useful to investigate special kinds of correlators like BPS ones in a supersymmetric theory.

\subsection{Direct reduction from affine $\slf(3)$ to W$_3$-algebra}
\label{sec:Direct}

Combining the correlator relations in subsections \ref{sec:BPW3} and \ref{sec:restricted}, we  can obtain a direct relation
between $\slf(3)$ current algebra and W$_3$-algebra.
However, restrictions on momenta have been put on the correlator of $\slf(3)$ WZNW model.
One may want to start from a generic correlator as in  subsection \ref{sec:previous}. 
The reduced action is given by \eqref{actionBP0}, which corresponds to the first realization of BP-algebra after performing a shift $\gamma^2 -1 \to \gamma^2$. 
In section \ref{sec:FreeBP}, we have developed a map of correlators described by two free field realizations.
Mapping to the second realization of BP-algebra, we can further reduce to $\slf (3)$ Toda field theory by applying the correlator relation in subsection \ref{sec:BPW3}.
However, twist operators $e^{\pm X}$ appearing in \eqref{tvertexsl3} would transform non-trivially under the shift $\gamma^2 -1 \to \gamma^2$. Because of this, the explicit form of correlator relations would be complicated.
In this subsection, we provide an approach of avoiding the insertions of extra twist operators.

As in  subsection \ref{sec:previous}, we begin with a generic correlator  of $\slf(3)$ WZNW model in the form \eqref{corr} with the action \eqref{actionsl3} and the path integral measure \eqref{measuresl3}.
Integrating out $(\gamma^1,\beta_1)$ and $(\gamma^3,\beta_3)$, we arrive at the action with interaction terms in \eqref{actionsl3p}.
Here we only make the change of field as
\begin{align}
 \phi_1 +  \frac{1}{b} \log |u_1 \mathcal{B}_1| ^2  \to \phi_1 \, ,
\end{align}
then the interaction terms become
\begin{align}
\frac{1}{2 \pi k} \int d^2 z \left[ e^{b \phi_1} (1 - \gamma^2 u_1^{-1} u_3 \mathcal{B}_1^{-1} \mathcal{B}_3 ) (1 -\bar \gamma^2 {\bar u_1}^{-1} \bar u_3 {\bar{\mathcal{B}}}_1^{-1} \bar{\mathcal{B}}_3) - e^{b \phi_2} \beta_2 \bar \beta_2 \right] \, . \label{newaction}
\end{align}
New operators are inserted at $z = y_1^n$.

Let us focus on the holomorphic part by decomposing $\phi_i (z, \bar z) $ as in \eqref{dec}.
The interaction terms \eqref{newaction} are then given by a linear combination of 
\begin{align}
&Q_1 = \int dz e^{b \varphi _1} \, , \\
&Q_2 = \int dz  e^{b \varphi_1} \gamma^2 u_1^{-1} u_3 \mathcal{B}_1^{-1} \mathcal{B}_3 \, , \label{Q2}\\
&Q_3 =\int dz  e^{b \varphi_2}  \beta_2  \, . 
\end{align}
The functions of $\mathcal{B}_1^{-1} \mathcal{B}_3 $ in \eqref{Q2} can be removed by a shift of $\varphi_1$, but this produces extra functions  $\mathcal{B}_1 \mathcal{B}_3^{-1} $ in $Q_1$.
In other words, we can remove functions in either $Q_1$ or $Q_2$ but not in both.
As examined in section \ref{sec:FreeBP}, a free field realization of BP-algebra uses the set of screening charges $(Q_1,Q_3)$ or $(Q_2,Q_3)$ (after removing the extra functions).

We may choose  $(Q_1,Q_3)$  as the set of screening charges and treat $Q_2$ perturbatively as
\begin{align}
e^{ - \lambda Q_2} = \sum_{p=0}^\infty \frac{(- \lambda)^p}{p!} (  Q_2)^p \, .
\end{align}
Rewriting $Q_2$ as
\begin{align}
Q_2 = \frac{1}{2 \pi i}
\int d zu_1^{-1} u_3 \mathcal{B}_1^{-1} \mathcal{B}_3 \oint d \mu _2 \mu_2 ^ {-2} e^{\mu_2 \gamma ^2 } e^{b \varphi_1} \, , 
\end{align}
the correlation functions can be put in the form studied in subsection \ref{sec:BPW3}.
Now we can apply the reduction procedure in subsection \ref{sec:BPW3} and express the correlation function in terms of $\slf(3)$ Toda field theory.
However,   an $N$-point function of $\slf(3)$ WZNW model is written as an infinite sum of correlators of $\slf(3)$ Toda field theory.

\section{Free field realizations of W-algebras from $\slf(4)$}
\label{sec:freesl4}

In sections \ref{sec:FreeBP} and \ref{sec:corrsl3}, we have examined correlators for theories with the symmetry of W-algebras associated to $\slf(3)$. In particular, it was important to choose a convenient free field realization of BP-algebra in order to obtain simpler expressions of correlator relations.
In this and the following sections, the analysis of W-algebras is extended from those associated with $\slf(3)$ to $\slf(4)$.
As explained in the introduction, one can construct three types of W-algebras with non-regular embeddings of $\slf(2)$ to $\slf(4)$ by Hamiltonian reduction.
These embeddings are known as subregular, rectangular, and minimal corresponding to the partitions $4 = 3 +1$, $4 = 2 + 2$, and $4 = 2 + 1+1$, respectively.
In this section, we express the generators of these non-regular W-algebras in terms of free fields by making use of screening charges in appendix \ref{sec:screening}.
With the explicit expressions, we can obtain maps of vertex operators and correlation functions among different free field realizations of the W-algebras.

For W-algebras associated with $\slf(4)$, we need three free bosons $\varphi_i$ $(i=1,2,3)$ with OPEs
\begin{align}
\varphi_i (z) \varphi_j (w) \sim - G_{ij} \log (z -w) \, , \label{bosonOPEsl4}
\end{align}
where we have introduced matrices
\begin{align}
G_{ij} = 
\begin{pmatrix}
 2 & -1 & 0 \\
-1 & 2 & -1 \\
0 & -1 & 2 
\end{pmatrix} \, , \quad
G^{ij} = 
\begin{pmatrix}
 3/4 & 1/2 & 1/4 \\
1/2 & 1 & 1/2 \\
1/4 & 1/2 & 3/4
\end{pmatrix} \label{Gmatrices} \, .
\end{align}
We also use $\ell$ sets of ghost systems $( \gamma^\alpha , \beta_\alpha ) $ satisfying 
\begin{align}
\gamma^\alpha (z) \beta_{\alpha '} (w) \sim \frac{\delta^\alpha_{~ \alpha '}}{z -w} \label{ghostOPEs}
\end{align}
where $\alpha , \alpha ' \in \{1,2, \ldots , \ell\}$. The number of ghost systems depends on the particular W-algebra we are considering.  In the case of $\slf(4)$, we set
\begin{align}\label{bsl4}
b = \frac{1}{\sqrt{k -4}} 
\end{align}
with $k$ the level of $\slf(4)$ current algebra.%
\footnote{Compared with the notation in appendix \ref{sec:screening}, we change $\kappa  \to  i b^{-1}$ and $k \to - k$, see also footnote \ref{-k} for $\slf (3)$. Furthermore, we use $ i \partial \varphi_j = \alpha_j$, which is suitable for Lagrangian description. }

\subsection{$W^{(2)}_4$-algebra}
\label{sec:W24}

The W-algebra with the subregular embedding of $\slf(2)$ into $\slf(n)$ is denoted as $W_n^{(2)}$. The properties of such subregular W-algebras were examined, for example, in \cite{Feigin:2004wb}. In this subsection, we set $n=4$.
The generators of the algebra include a spin-one current $H(z)$, three spin-two currents $G^\pm(z)$, $T(z)$, and a spin-three current $W(z)$. Here $T(z)$ is the energy-momentum tensor with central charge
\begin{align}
c = 24 k+\frac{60}{k-4}-19 \, .
\end{align}
The OPEs involving $H(z)$ are
\begin{align}\label{OPEsHHHG}
H(z) H(w) \sim \frac{( 8 - 3 k)/4}{(z - w)^2} \, , \quad
H(z) G^\pm (w) \sim \pm  \frac{G^+ (w)}{z-w} \, .
\end{align}
The spin-three current $W(z)$ is regular with respect to $H(z)$. One may find expressions of other OPEs in \cite{Feigin:2004wb}.
The W$_4^{(2)}$ algebra can be realised by a set of ghost system $(\gamma,\beta)$ along with 
three free bosons $\varphi_i$ $(i=1,2,3)$. 

\bigskip

\noindent {\bf The first realization.}
The screening charges for $W^{(2)}_4$-algebra can be found in \eqref{subs}.
In one of the realizations,  these are given by screening operators
\begin{align} \label{screeningW241}
\mathcal{V}_1 = e^{ b \varphi_1 }  \beta  \, , \quad \mathcal{V}_2 =e^{ b \varphi_2 }   \gamma  \, , \quad
\mathcal{V}_3 = e^{  b \varphi_3 } \, .
\end{align}
Requiring generators to commute with the screening charges, we find the expressions of $T$ and $H$ as
\begin{align}
T = - \frac12 G^{ij} \partial \varphi_i \partial \varphi_j + \frac{(2k -  5)b}{2}  \partial^2 \varphi_1 + (k-2)b   \partial^2 \varphi_2 +    \frac{(2k -  5)b}{2}   \partial^2 \varphi_3  + \gamma \partial \beta
\end{align}
and
\begin{align}
H = - \frac{1}{4 b} \partial \varphi_1 + \frac{1}{2 b} \partial \varphi_2 + \frac{1}{4 b} \partial \varphi_3 - \gamma \beta \, .
\end{align}
In particular, the conformal weights of $(\gamma ,\beta)$ are $(1,0)$. 
We also find that
\begin{align}
\begin{aligned}
&G^+ = - \gamma \gamma \beta - b^{-1} \partial \varphi_1 \gamma + (k -2) \partial \gamma   \, , \\
&G^- = - b^{-1} \left(2 (k-2)  \partial \varphi_2 \partial \beta +(k-2)  \partial \varphi_3 \partial \beta  +2 \partial \varphi_2 \gamma \beta \beta  + \partial \varphi_3 \gamma \beta \beta \right) \\
& \quad - (k-3)  b^{-1} \partial^2 \varphi_2 \beta +k \partial \varphi_2 \partial \varphi_2 \beta   +k \partial \varphi_3 \partial \varphi_2 \beta  -4 \partial \varphi_2 \partial \varphi_2 \beta  -4 \partial \varphi_3 \partial \varphi_2 \beta \\
& \quad +3 k \gamma \beta \partial \beta +k \partial \gamma \beta \beta -6 \gamma \beta \partial \beta -4 \partial \gamma \beta \beta  + \gamma \gamma \beta \beta \beta + \left(5 + k^2 - 9 k/2 \right) \partial \partial \beta 
\end{aligned}
\end{align}commute with the screening charges, and they indeed satisfy the correct OPEs with $T$ and $H$.
The expression of $W$ can then be found from the OPE with $G^+$ and $G^-$ as 
\begin{align}
W = \frac{4 (k-4) (5 k-16)}{3 (8-3 k)^2 (k-2)} \gamma \gamma \gamma \beta \beta \beta + \cdots \, ,
\end{align}
which commutes with the screening charge.

\bigskip

\noindent {\bf The second realization.}
Another set of screening charges is given by
\begin{align}  \label{screeningW242}
\mathcal{V}_1 = e^{ b \varphi_1 } \beta  \, , \quad \mathcal{V}_2 = e ^{ b \varphi_2 } \, , \quad
\mathcal{V}_3 = e^{  b \varphi_3 } \, .
\end{align}
Using the same commuting condition, we find the expressions of $T$ and $H$ as
\begin{align}
T = - \frac12 G^{ij} \partial \varphi_i \partial \varphi_j + \frac{3b}{2} \partial^2 \varphi_1 + (k-2)b \partial^2 \varphi_2 +  \frac{(2k -  5)b}{2}  \partial^2 \varphi_3  -\gamma \partial \beta  - 2 \partial \gamma \beta
\end{align}
and
\begin{align}
H = \frac{3}{4 b} \partial \varphi_1 + \frac{1}{2 b} \partial \varphi_2 + \frac{1}{4 b} \partial \varphi_3 + \gamma \beta \, .
\end{align}
In particular, the conformal weights of $(\gamma,\beta)$ are now $(-1,2)$. 
Using OPEs, such as \eqref{OPEsHHHG}, as constraints, we found
\begin{align}
G^+ = \beta \, ,\quad G^- = - \gamma \gamma \gamma \gamma \beta \beta \beta + \cdots \, , \quad
W = -\frac{4 (k-4) (5 k-16)}{3 (8-3 k)^2 (k-2)} \gamma \gamma \gamma \beta \beta \beta + \cdots \, ,
\end{align}
which also commute the screening charges. \\

As in \eqref{twist}, we then twist the energy-momentum tensor as
\begin{align}
T(z) \to T_t (z) = T(z) + \partial H(z) \, . \label{twistW24}
\end{align}
The spins of $G^+$ and $G^-$ become $1$ and $3$, respectively, with respect to the twisted energy-momentum tensor.
For the both free field realizations, the conformal weights of $(\gamma, \beta)$ are $(0,1)$ after the twisting.

\subsection{Rectangular W-algebra with $\mathfrak{su}(2)$ symmetry}
\label{sec:rectangular}

A generic rectangular W-algebra is obtained by performing Hamiltonian reduction of Lie algebra $\slf(Mn)$. The embedding of $\slf(2)$ into $\slf(Mn)$ corresponds to the partition $Mn=n+n+\ldots+n$, a sum of $M$ numbers of $n$. A rectangular W-algebra contains a subalgebra $\slf(M)$.
The rectangular W-algebra can be realized as the asymptotic symmetry of extended higher spin gravity  \cite{Creutzig:2018pts,Creutzig:2019qos}.%
\footnote{See \cite{Creutzig:2013tja,Eberhardt:2018plx,Creutzig:2019wfe,Rapcak:2019wzw,Eberhardt:2019xmf} for related works.}
In this subsection, we examine free field realizations of the rectangular W-algebra associated to $\slf(Mn)$  with $M=n=2$.

The rectangular W-algebra associated with $\slf(4)$ is generated by three spin-one currents $J^a(z)$  and four spin-two currents $T(z),Q^a(z)$, where $a=1,2,3$ and $T(z)$ being the energy-momentum tensor with central charge
\begin{align}
c = 12 k+\frac{60}{k-4}+7 \, .
\end{align}
The spin-one currents $J^a(z)$ generate an $\mathfrak{su}(2)$ current algebra with OPEs\begin{align}
\begin{aligned}
&J^a (z) J^b (w) \sim \frac{{\ell} \delta^{a,b}/2}{(z-w)^2} + \frac{i f^{ab}_{~~c} J^c(w)}{z-w} \, ,\\
&T(z) J^a (w) \sim \frac{J^a(w)}{(z-w)^2} + \frac{\partial J^a (w)}{z-w} \, ,
\end{aligned}
\end{align}
where $\ell = 4 -2 k$ is the $\mathfrak{su}(2)$ level. The structure constant of $\mathfrak{su}(2)$ is given by $i f^{abc}$ whose indices can be raised and lowered by $\kappa^{ab} = \delta^{a,b}/2$ and $\kappa_{ab} = 2 \delta_{a,b}$,
.
The spin-two currents $Q^a(z)$ transform as in the adjoint representation of $\mathfrak{su}(2)$.
Their OPEs with  $J^a(z)$ and $T(z)$ are
\begin{align}
J^a (z) Q^b (w) \sim \frac{i f^{ab}_{~~c} Q^c (w)}{z-w} \, , \quad
T(z) Q^b (w) \sim \frac{2 Q^b (w)}{(z-w)^2} + \frac{\partial Q^b (w)}{z-w} \, .
\end{align}
The OPE $Q^a(z) Q^b (w)$ can be found in \cite{Creutzig:2018pts,Creutzig:2019qos}.
The free field realization of the theory can be constructed from two sets of ghost systems $(\gamma^\alpha , \beta_\alpha )$ $(\alpha =1,2)$ along with three free bosons $\varphi_i$ $(i=1,2,3)$.

\bigskip

\noindent {\bf The first realization.}
As found in \eqref{i2}, there are two sets of screening charges for the rectangular W-algebra of $\slf(4)$.
One of them is given by the set of screening operators
\begin{align}
\mathcal{V}_1 = e^{ b \varphi_1 } \beta_1  \, , \quad \mathcal{V}_2 =e^{ b \varphi_2 }  (\gamma^1 - \gamma^2)  \, , \quad
\mathcal{V}_3 =e^{ b \varphi_3 }  \beta_2  \, .
\end{align}
The energy-momentum tensor commuting with the screening charges is found to be 
\begin{align}
T = - \frac12 G^{ij} \partial \varphi_i \partial \varphi_j  +  \frac{(k-1)b}{2 } \partial^2 \varphi_1  +  (k-2)b  \partial^2 \varphi_2 + \frac{(k-1) b}{2 }  \partial^2 \varphi_3 -  \partial \gamma^1 \beta_1-  \partial \gamma^2 \beta_2 \label{recEM}.
\end{align}
The $\mathfrak{su}(2)$ currents in terms of the free fields are given by
\begin{align}
\begin{aligned}
&J^+ = \beta_1 + \beta_2 \, , \quad J^3 = \gamma^1 \beta _1 + \gamma ^2 \beta _2 +\frac{1}{2 b}  \partial \varphi_1 +\frac{1}{2 b} \partial \varphi_3 \, , \\
&J^- = - b^{-1} \partial \varphi_3 \gamma ^2 - b^{-1} \partial \varphi_1 \gamma ^1 - \gamma ^1 \gamma ^1 \beta _1 - \gamma ^2 \gamma ^2 \beta _2 +(k-2) \partial  \gamma ^1 +(k-2) \partial  \gamma ^2 \, ,
\end{aligned}
\end{align}
where we have defined $J^\pm = J^1 \pm i J^2$.
A charged spin-two current is found to be
\begin{align}
\begin{aligned}
&Q^+ = \partial \varphi _1 \beta _1 + \partial \varphi_1 \beta _2 +2 \partial \varphi_2 \beta _1 -2  \partial \varphi_2 \beta_2 - \partial \varphi_3 \beta _1 - \partial \varphi_3 \beta _2 \\
& \qquad+ 4 b \gamma ^1 \beta_1 \beta_2  -  4 b \gamma ^2  \beta _1 \beta _2 + 2 (k-2) b \partial \beta _2  - 2 (k-2) b \partial \beta _1
\end{aligned}
\end{align}
with $Q^+ = Q^1 + i Q^2$. The rest of charged spin-two currents can be obtained from their OPEs with $J^-$.

\bigskip

\noindent {\bf The second realization.}
The other set of screening operators presented in \eqref{i2} is given by
\begin{align}
\mathcal{V}_1 = e^{ b \varphi_1 } \beta_1  \, , \quad \mathcal{V}_2 = e^{ b \varphi_2 } (1 - \gamma^1  \gamma^2)  \, , \quad
\mathcal{V}_3 = e^{  b \varphi_3 } \beta_2  \, .
\end{align}
The energy-momentum tensor has exactly the same expression as in the previous realization as in \eqref{recEM}.
The $\mathfrak{su}(2)$ currents, in this case, are
\begin{align}
\begin{aligned}
&J^+ = - b^{-1} \partial \varphi_3 \gamma ^2 - \gamma ^2 \gamma ^2 \beta _2 +\beta _1 +(k-2) \partial  \gamma ^2 \, , \\
&J^3 = \gamma^1  \beta _1 - \gamma ^2 \beta _2 -\frac{1}{2b} \partial \varphi_3  +\frac{1}{2 b} \partial \varphi_1  \, , \\
&J^- = - b^{-1} \partial \varphi_1 \gamma ^1 - \gamma ^1 \gamma ^1 \beta _1 +\beta _2 +(k-2) \partial  \gamma ^1 \, .
\end{aligned}
\end{align}
A charged spin-two current is found to be
\begin{align}
\begin{aligned}
&Q^3 = 
- \gamma^1 \gamma ^1  \gamma ^2 \gamma ^2 \beta _1 \beta _2  + \cdots 
\end{aligned}
\end{align}
up to an overall normalization.  The rest of the charged spin-two currents can again be obtained by their OPEs with $J^\pm$. \\

Note that we do not twist the energy-momentum tensor here.
In fact, the conformal dimensions of $(\gamma^\alpha , \beta_\alpha )$ $(\alpha =1,2)$ are $(0,1)$ with respect to the energy-momentum tensor \eqref{recEM}.

\subsection{QSCA with $\mathfrak{su}(2)$ symmetry}
\label{sec:QSCA}

The W-algebra with minimal embedding of $\slf(2)$ into $\slf(n)$ was studied in \cite{Romans:1990ta}.
The algebra is known as quasi-superconformal algebra (QSCA), since it can be obtained by replacing fermionic spin-3/2 generators of the superconformal algebras presented in \cite{Knizhnik:1986wc,Bershadsky:1986ms} with bosonic generators of the same spin.
In this subsection, we examine free field realizations of QSCA associated to $\slf(4)$. 

The generators  of QSCA include four spin-one currents $H(z),J^a(z)$ $(a=1,2,3)$, four spin-$3/2$ bosonic currents $G_\pm^{i} (z)$ $(i=1,2)$, and a spin-two current $T (z)$, the energy-momentum tensor with central charge
\begin{align}
c =6 k-\frac{60}{4-k}+15 \, .
\end{align}
The OPEs involving $H(z)$ satisfy
\begin{align}
H (z) H(w) \sim \frac{1}{(z-w)^2} \, , \quad H (z) G_{\pm}^i (w) \sim \pm \frac{q G_\pm^i (w)}{z - w} 
\end{align}
with $q = 1/\sqrt{2-k}$. The rest of the spin-one currents $J^a(z)$ generate an $\mathfrak{su}(2)$ current algebra. The currents are normalised to satisfy the OPE
\begin{align}
J^a (z) J^b (w) \sim \frac{\ell /2}{(z - w)^2} + \frac{i f^{ab}_{~~c} J^c (w) } {z-w}  
\end{align}
with the level $\ell  = 1 - k$. The action of $J^a(z)$ on $G^i_\pm(w)$ is given by 
\begin{align}
J^a (z) G_+^i (w)\sim - \frac{(\sigma^a)^i _{~ j} G^j_+ (w)}{z-w} \, , \quad
J^a (z) G_-^{i} (w) \sim \frac{G_-^{j} (w)(\sigma^a)_{j}^{~i}  }{z-w} \, .
\end{align}
To compute the coefficients of the OPEs, we first define the following matrices
\begin{align}
\sigma^1_{ij} = \frac12 
\begin{pmatrix}
0 & 1 \\
1 & 0 
\end{pmatrix} \, ,\quad
\sigma^2_{ij}= \frac12 
\begin{pmatrix}
0 & - i \\
i & 0 
\end{pmatrix} \, ,\quad
\sigma^3_{ij} = \frac12 
\begin{pmatrix}
1 & 0 \\
0 & -1
\end{pmatrix} 
\end{align}
whose indices can then be raised to the form of coefficients by the application of $\delta^{i,j}$.
The non-trivial OPEs among the spin-$3/2$ generators $G_\pm^i$ are
\begin{align}
\begin{aligned}
&G_+^i (z) G_-^j (w) \sim \delta^{i,j} \left[ \frac{c_1 }{(z-w)^3} + \frac{2 c_2 H (z)}{(z-w)^2} + \frac{c_2 \partial H(w) -2 T(w) + c_4 (HH)(w)}{z-w}\right] \\
& \quad + (\sigma^{a})^{i j}   \left[ \frac{2 c_5 J^a (w)}{(z-w)^2} + \frac{c_5 \partial J^a (w) + c_6 (H J^a) (w)}{z-w}\right] 
+ c_7 \delta^{i,j} \delta^{a,b}\frac{(J^a J^b) + (J^b J^a)}{z-w}
\end{aligned}
\end{align}
with 
\begin{align}
\begin{aligned}
&c_1 =  -\frac{4 (k-2) (k-1)}{k-4} \, , \quad c_2 =  \frac{2 \sqrt{2-k} (k-1)}{k-4}\, ,  \quad c_4 = \frac{6}{k-4}+3 \, , \\ &c_5 =  -\frac{8}{k-4}-4 \, , \quad c_6 =  \frac{8 \sqrt{2-k}}{k-4} \, , \quad c_7 =  -\frac{2}{k-4} \, .
\end{aligned}
\end{align}
The generators of the QSCA can be expressed in terms of three ghost systems $( \gamma^\alpha , \beta_\alpha)$ $(\alpha =1,2,3)$ and three free bosons $\varphi_i$ $(i=1,2,3)$.

\bigskip

\noindent {\bf The first realization.}
Sets of screening charges for QSCA can be found in \eqref{QSCAi} and \eqref{QSCAh}.
One set of screening operators  in \eqref{QSCAi} is given by
\begin{align} \label{screeningmin1}
\mathcal{V}_1 =e^ { b \varphi_1 }   \beta_1  \, , \quad \mathcal{V}_2 = e^{ b \varphi_2 }  (\beta_2 - \gamma^1 \beta_3)  \, , \quad
\mathcal{V}_3 =  e^{  b \varphi_3 } 
\end{align}
from which we find the expression of the energy-momentum tensor to be
\begin{align}
T = - \frac12 G^{ij} \partial \varphi_ i \partial \varphi_j + Q^i \partial^2 \varphi_i  + \sum_{\alpha=1}^3 \left( a_\alpha \gamma^\alpha  \partial \beta_\alpha + (a_\alpha -1) \partial \gamma^\alpha \beta_\alpha  \right) \label{minEM}
\end{align}
where
\begin{align}
Q^1 = \frac{3 b}{2 } \, , \quad Q^2 = 2 b  \, , \quad Q^3 = \frac{(k-1) b}{2 } \, , \quad a_1 = 0 \, , \quad a_2 = a_3 = - \frac12.
\end{align}
The conformal weights of $(\gamma^\alpha , \beta_\alpha)$ are $(a_\alpha , 1 - a_\alpha)$.
The  $\mathfrak{su}(2)$ currents commuting with the screening charges are written as
\begin{align}
J^3 = j^3 + \tilde \jmath^3 \, , \quad J^\pm = j^\pm + \tilde \jmath^\pm 
\end{align}
with
\begin{align}
\begin{aligned}
&j^+ = \beta_1 \, , \quad j^3 = \gamma^1 \beta_1 + \frac{1}{2 b} \partial \varphi_1 \, , \quad
j^- = - \frac{1}{b} \partial \varphi_1 \gamma^1 - \gamma^1 \gamma^1 \beta_1 +(k-2) \partial \gamma^1 \, , \\
&\tilde \jmath^+ = - \gamma^2 \beta_3 \, , \quad \tilde \jmath^3 = - \frac12 (\gamma^2 \beta_2 - \gamma^3 \beta_3) \, , \quad \tilde \jmath^- = - \gamma^3 \beta_2 \, .
\end{aligned}
\end{align}
The $\mathfrak{u}(1)$ current in terms of the free fields is given by
\begin{align}
H = q \left( b^{-1} \partial \varphi^2 + \gamma^2 \beta_2 + \gamma^3 \beta_3 \right) \, , 
\end{align}
where $q=1/\sqrt{2-k}$. The spin-3/2 currents are found to be
\begin{align}
\begin{aligned}
&G_+^1 = \beta_2 \, ,\quad G_+^2 = \beta_3 \, , \\
&G_-^1 = \frac{2}{k -4} \gamma^1 \gamma^1 \gamma^2 \gamma^2 \beta_1 \beta_3 + \cdots \, , \quad
G_-^2 =   \frac{2}{k -4} \gamma^1 \gamma^1 \gamma^2 \gamma^2 \beta_1 \beta_2 + \cdots \, .
\end{aligned}
\end{align}

\bigskip

\noindent {\bf The second realization.}
A second set of screening operators  in \eqref{QSCAi} is
\begin{align} \label{screeningmin2}
\mathcal{V}_1 =e^{ b \varphi_1 }   \beta_1  \, , \quad \mathcal{V}_2 =e^{ b \varphi_2 }  (\beta_2 - \gamma^1 \beta_3)  \, , \quad
\mathcal{V}_3 = e^{  b \varphi_3 }  \gamma^2  \, . 
\end{align}
The energy-momentum tensor commuting with the screening charges takes the same form as \eqref{minEM}, but with a different set of parameters:
\begin{align}
\begin{aligned}
Q^1 =  \frac{3 b}{2}\, , \quad Q^2 =  \frac{k b }{2 } \, , \quad Q^3 =  \frac{( k-1 ) b}{2 } \, , \quad
a_1 = - a_2 = - \frac12 \, , \quad a_3 =0 \, .
\end{aligned}
\end{align}
The spin-one currents are
\begin{align}
\begin{aligned}
&J^+ = \beta_3 \, , \quad
J^3 = \frac{1}{2} \gamma^1 \beta_1 +\frac{1}{2} \gamma^2  \beta_2 + \gamma^3 \beta_3 +\frac{1}{2 b } \partial \varphi_1 +\frac{1}{2 b } \partial \varphi_2  \, , \\
&J^- = - b^{-1} \partial \varphi_1 \gamma^3  - b^{-1} \partial \varphi_2 \gamma^3  - b^{-1}\partial \varphi_1 \gamma^1 \gamma^2  +(k-2)  \partial \gamma^1 \gamma^2  - \gamma^1 \gamma^1 \gamma^2 \beta_1  \\
& \quad \quad - \gamma^1 \gamma^3  \beta_1  - \gamma^2  \gamma^3 \beta_2  - \gamma^3  \gamma^3  \beta_3 +(k-1) \partial \gamma^3  \, , \\
&H = q \left ( \gamma^1 \beta_1 - \gamma^2 \beta_2  +  \frac{1}{2 b } \partial \varphi_1  + \frac{1}{2 b}  \partial \varphi_3 \right) \, .
\end{aligned}
\end{align}
Two of the spin-$3/2$ currents are
\begin{align}
\begin{aligned}
&G_+^2 = \beta_1 - \beta_3 \gamma^2 \, , \\
&G_-^1 = - 2 b \partial \varphi_1 \gamma^1 \beta _3  - 2 b \partial \varphi_3 \beta_2  + \left(\frac{4}{k-4}+2\right)  \partial \gamma^1 \beta_3 + \frac{2}{k-4} \gamma^2 \beta_2  \beta _2 \\
& \quad \quad  - \frac{2}{k-4}  \gamma^1  \gamma^1  \beta_1 \beta _3  + \left(\frac{4}{k-4}+2\right) \partial \beta_2 \, .\\
\end{aligned}
\end{align}
The expressions of the other two spin-$ 3 / 2$ currents $G_+^1,G_-^2$ can be found from their OPEs with $J^-$.

\bigskip

\noindent {\bf The third realization.}
The last set of screening operators presented in \eqref{QSCAi} is given by
\begin{align} \label{screeningmin3}
\mathcal{V}_1 = e^{b \varphi_1 } \beta_1  \, , \quad \mathcal{V}_2 = e^{ b \varphi_2 } (\beta_2 - \gamma^1 \beta_3)  \, , \quad
\mathcal{V}_3 = e^{  b \varphi_3 }   \gamma^3 \, .
\end{align}
The energy-momentum tensor  is \eqref{minEM} with
\begin{align}
\begin{aligned}
& Q^1 =  \frac{(k-1) b}{2 } \, , \quad Q^2 =  \frac{k b }{2 } \, , \quad Q^3 = \frac{( k-1 ) b }{2 } \, , 
&a_1 = a_3 = \frac12 \, , \quad a_2 = 0 
\end{aligned}
\end{align}
and the spin-one currents are
\begin{align}
\begin{aligned}
&J^+ = \beta_2 \, , \quad
J^3 =  -\frac{1}{2} \gamma^1  \beta _1 + \gamma ^2 \beta _2 +\frac{1}{2} \gamma ^3  \beta _3 +\frac{1}{2 b} \partial \varphi_2  \, , \\
&J^- = - b^{-1} \partial \varphi _2  \gamma^2 + \gamma^1 \gamma^2  \beta _1 - \gamma^2  \gamma^2  \beta_2 - \gamma^2  \gamma ^3 \beta _3 + \gamma^3 \beta_1 +(k-1) \partial  \gamma ^ 2 \, , \\
&H = - q \left(\gamma^1 \beta _1 + \gamma ^3 \beta _3 - \frac{1}{2 b}\partial \varphi_3 + \frac{1}{2 b} \partial \varphi_1 \right) \, .
\end{aligned}
\end{align}
Two of the spin-$3/2$ currents are found to be
\begin{align}
\begin{aligned}
&G_+^2 = \partial \varphi_1 \gamma^1 + b \gamma^1 \gamma^1 \beta_1 + b \gamma^3 \beta_2 + (2-k) b \partial \gamma^1 \, ,  \\
&G_-^1 =2  \partial \varphi_3 \beta_3 - 2 b  \gamma ^3  \beta _ 3 \beta _3  - 2 b  \beta_1 \beta_2 - 2 (k-2) b \partial \beta _3  \, , \\
\end{aligned}
\end{align}
and the other two $G_+^1,G_-^2$ can be obtained by the action of $J^-$. 

\bigskip

For the above three types of free field realizations, we consider the twist of energy-momentum tensor by
\begin{align}
T(z) \to T_t(z) = T(z) + \frac{1}{2 q} \partial H \, , \label{twistmin}
\end{align}
where $q=1/\sqrt{2-k}$. With respect to the twisted energy-momentum tensor, the conformal weights of $G_+^i$ and $G_-^i$ become 1 and 2, respectively.
Whereas, all three pairs of ghosts $(\gamma^\alpha , \beta_\alpha )$ with $\alpha =1,2,3$ have conformal weights $(0,1)$ as desired.

\bigskip

\noindent {\bf The fourth realization.}
Finally, let us consider the free field realization associated with the set of screening operators in \eqref{QSCAh}:
\begin{align} \label{screeningsmin4}
\mathcal{V}_1 =e^{ b \varphi_1 }   (\beta_1 + \gamma^2 \beta_3 )  \, , \quad
\mathcal{V}_2 =  e^ { b \varphi_2 }  \beta_2  \, , \quad
\mathcal{V}_3 = e^{  b \varphi_3 }  ( \gamma^3 - \gamma^2 \gamma^1 ) \, . 
\end{align}
The energy-momentum tensor is again given by \eqref{minEM} with
\begin{align}
\begin{aligned}
& Q^1 =  \frac{(k-1) b}{2 } \, , \quad Q^2 =  \frac{k b }{2 } \, , \quad Q^3 = \frac{( k-1 ) b }{2 } \, , 
&a_1 =a_3 =  \frac12 \, , \quad a_2 = 0 \, .
\end{aligned}
\end{align}
The spin-one currents are
\begin{align}
\begin{aligned}
&J^+ = \beta_2  + \gamma^1 \beta_3 \, , \quad
J^3 =  - \frac{1}{2} \gamma ^1 \beta _1 + \gamma ^2 \beta _2  + \frac{1}{2} \gamma ^3 \beta _3 +\frac{1}{2 b} \partial \varphi_2   \, , \\
&J^- =- b^{-1} \partial \varphi_2 \gamma^2 + \gamma ^3 \beta _1 - \gamma ^2 \gamma ^2 \beta _2 +(k-2) \partial \gamma ^2  \, , \\
&H =- q \left( \gamma ^1 \beta _1 + \gamma ^3 \beta _3 + \frac{1}{2 b}\partial \varphi_1  - \frac{1}{2 b} \partial \varphi_3 \right)  \, .
\end{aligned}
\end{align}
Two of the spin-$3/2$ currents are
\begin{align}
\begin{aligned}
&G_+^2 =  \partial \varphi_1 \gamma ^ 1 +  b \gamma ^1 \gamma ^1 \beta _1  - b \gamma ^ 1 \gamma ^2 \beta ^2  + b \gamma ^ 1 \gamma ^3 \beta _3 + b \gamma ^ 3 \beta _2  + (1-k) b \partial \gamma ^ 1  \, ,  \\
&G_-^1 =  2  \partial \varphi_3 \beta _3 - 2 b \gamma ^ 1 \beta _1 \beta _3  -  2 b  \gamma ^ 2  \beta _2  \beta _3  - 2 b \gamma ^ 3 \beta _3 \beta _3  - 2 b \beta _1 \beta _2  - 2 (k-1) b \partial  \beta _3  \, , \\
\end{aligned}
\end{align}
and rest of the spin-$3 / 2$ currents can again be obtained by the action of $J^-$ on them.
These generators are consistent with those in \cite{Romans:1990ta}.

In this realization, we may twist the energy-momentum tensor by
\begin{align}
T(z) \to T_t(z) = T(z) +a_1 \partial J^3  +  a_2 \partial H   \, .
\end{align}
However, unlike the previous realizations, this twist does not allow the conformal dimensions of $(\gamma^\alpha , \beta_\alpha )$ for all $\alpha =1,2,3$ to be $(0,1)$.
Even though other values of conformal dimensions are also permitted, it introduces difficulties to construct maps from this free field realization to the other three.

\subsection{Vertex operators}

So far, we have constructed generators for three different non-regular W-algebras associated to $\slf(4)$ in terms of free fields. In order to construct vertex operators for each W-algebra, we may twist the energy-momentum tensor such that the conformal weights of $(\gamma^\alpha , \beta_\alpha)$ become $(0,1)$ as we did for W$^{(2)}_4$-algebra in \eqref{twistW24} and for the first three QSCA realizations in \eqref{twistmin}. Such twists are not needed for the rectangular case since the desirable ghost conformal dimensions are achieved with the original energy-momentum tensor. 
As discussed at the end of section \ref{sec:QSCA}, it is not possible to find a good twist for the fourth realization of QSCA. It is, therefore, difficult to establish a correspondence between this realization with the others. We shall now disregard this exceptional case, and instead, we examine rest of the realizations of these three non-regular W-algebras in general.

For any of the three algebras, its generators are denoted by $F^a$, where $a$ runs from 1 to the dimension of algebra.
The conformal dimension of $F^a$ with respect to the (twisted) energy-momentum tensor is denoted by $h^a$.
Since $h^a$ is integer for all $a$ in all W-algebras,
the generators can be expanded as
\begin{align}
F^a (z) = \sum_{n \in\mathbb{Z}} \frac{F^a_n}{z^{n + h^a}}.
\end{align}
A vertex operator and its corresponding primary state is define as
\begin{align}
 \label{mubasis}
V_{ \ell^i } ( \mu_\alpha | z) = e^ { \mu_\alpha \gamma^\alpha } e^{  \ell^i \varphi_i } \, , \quad
| \ell^i ; \mu_\alpha \rangle \equiv \lim_{z \to 0} V_{ \ell^i }(\mu_\alpha | z) | 0 \rangle \, .
\end{align}
The modes of the generators act on the primary states as 
\begin{align}
\begin{aligned}
&F^a_n |  \ell^i  ; \mu_\alpha \rangle= 0 , \quad
&F^a_0 | \ell^i  ; \mu_\alpha \rangle= - \mathcal{D}_{F^a}  |  \ell^i  ; \mu_\alpha \rangle \, ,
\end{aligned}
\end{align}
where $n > 0$ and $\mathcal{D}_{F^a}$ are differential operators with respect to $\mu_\alpha$.

The explicit form of $\mathcal{D}_{F^a}$ can be found from the free field expressions of the generators.
These differential operators therefore have different expressions for different  realizations.
However, it can be shown that the differential operators $\mathcal{D}_{F^a}$ satisfy the same commutation relations  for all free field realizations of the same W-algebra.
The second-, third-, and fourth-order Casimir operators $\mathcal{D}_p$ with $p=2,3,4$ which commute with $D_{F^a}$ are constructed. We shall avoid presenting these tedious expressions in this paper, since they are not directly related to the calculation. These Casimir operators are linear combinations of terms of the form
$
\mathcal{D}_{F_{a_1}} \cdots \mathcal{D}_{F_{a_n}} 
$
with $n \leq p$.
In particular, the eigenvalue with respect to the second-order Casimir operator $\mathcal{D}_2$ corresponds to the conformal weight in an analogous way as in \eqref{cw} for the BP-algebra case.
The eigenvalues of these Casimir operators on the primary states defined in \eqref{mubasis} are the same for all free field realizations of the same W-algebra.

It is desirable to obtain maps of correlation functions (or vertex operators) among different free field realizations for a non-regular W-algebra. The form of vertex operators defined in \eqref{mubasis} are not suitable for this purpose. The map can be realised  more clearly if we make a change of basis such that the vertex operators take the form $P(\gamma^\alpha ) \exp ( \ell^i \varphi_i ) $, where $P(x ) $ is a polynomial of $x$.
A convenient choice for making such a change of basis is
\begin{align}
V_{ \ell^i } (a_1 , \ldots ,  a_n) = \int \frac{d \mu}{\mu} \mathcal{D}_{a_1} \cdots \mathcal{D}_{a_n} V_{ \ell^i } (\mu | z) \, .
\end{align}
This is a generalisation of the change to $m$-basis we used in subsection \ref{sec:maps} for the BP-algebra. 
In the case of W$_4^{(2)}$-algebra, the basis is essentially the same as for the BP-algebra, and therefore the arguments in subsection \ref{sec:maps} can be directly applied.
For other non-regular W-algebras, it is also possible to find the maps of vertex operators explicitly at least for concrete examples.

\section{Correlator relations for W-algebras from $\slf(4)$}
\label{sec:corrsl4}

Using the screening charges in appendix \ref{sec:screening}, we have developed free field realizations of non-regular W-algebras associated to $\slf(4)$. This allows us to derive new correspondences among correlation functions involving these W-algebras. There are five different partitions of the integer $4$ which corresponds to five different W-algebras.
Recall that a similar relation was found between the $\slf(2)$ WZNW model with partition $2 = 1+1$ and the Virasoro algebra (or W$_2$-algebra) with partition $2 =2$.
Moreover, for the $\slf(3)$ case, we have obtained the reduction relations for algebras with partitions from $3=1+1+1$ to $3 = 2+1$ and from $3=2+1$ to $3=3$.
Observing the pattern, one can deduce that, in the $\slf(4)$ case, correlator relations can be obtained directly using the generic procedure for W-algebras with the following partitions
\begin{align}
\begin{aligned}
4 &=1+1+1+1 &\to&  &4&=2+1+1 \, , \\
4&=2+1+1 &\to & &4&=3+1 \, ,  \\  
4&=3+1 &\to& &4&=4 \, .
\end{aligned}
\end{align}
In this section, we will study these cases in detail.

Other than the above cases, there are many relations we can obtain for W-algebras associated to $\slf (4)$.
For instance, we can perform a direct reduction to obtain correlator relation for $4=2+1+1 \to 4 = 2+2$.
In section \ref{sec:corrsl3}, we have discussed how to obtain direct relations for $3=1+1+1 \to 3=3$, and
 analogous techniques can be applied to the $\slf(4)$ case as well. 
 In particular, it is possible to relate $\slf(4)$ current algebra to all free field realizations of W-algebras by putting restrictions on momenta of vertex operators. 
This kind of reductions may be regarded as correlator versions of what have been done in appendix \ref{sec:screening} to obtain screening charges for various W-algebras. An application of the reduction from $\slf(4)$ current algebra to rectangular W-algebra  can be found in \cite{Creutzig:2018pts,Creutzig:2019qos}.

\subsection{Reduction from affine $\slf(4)$ to QSCA}

In this subsection, we reduce the $\slf(4)$ WZNW model to a free field realization of QSCA with $\mathfrak{su}(2)$ subalgebra in it. 
This reduction relation was already obtained in \cite{Creutzig:2015hla} where the reduced theory corresponds to the fourth realization of QSCA with the screening operators \eqref{screeningsmin4}.
As observed in the previous section, it is not easy to construct maps from the fourth realization of QSCA to the others. We therefore consider the reduction to the third realization instead in this subsection. 
We can then create maps among the three QSCA realizations and reduce the QSCA theory further to the subregular or the regular W-algebras.

We start from a first order formulation of the $\slf(4)$ WZNW model.
The analysis in appendix \ref{sec:screening} shows that it is important to choose a proper expression of the action:
\begin{align}
\begin{aligned}
S &= \frac 1 {2\pi}\int \dd^2 z\Big[\frac{G^{ij}}{2}\p\phi_{i}\pb\phi_{j}+\frac{b}{4}\sqrt{g}\mathcal{R}(\phi_1+\phi_2+\phi_3)+\sum_{\alpha=1}^6(\beta_\alpha \pb\gamma^\alpha+\bb{\alpha} \partial\gb{\alpha})\Big]\\
& \quad - \frac 1 {2\pi k}\int \dd^2 z\Big[e^{b\phi_{1}}\beta_1\bb{1}+e^{b\phi_{2}}(\beta_2-\gamma^1\beta_4)(\bb{2}-\gb{1}\bb{4})\\
&\hspace{3cm}+e^{b\phi_{3}}(\beta_3-\gamma^2\beta_5-\gamma^4\beta_6)(\bb{3}-\gb{2}\bb{5}-\gb{4}\bb{6})\Big] \, ,
\end{aligned} \label{actionsl4}
\end{align}
which corresponds to the free field realization of $\slf(4)$ current algebra with screening charges \eqref{sl4min}.
The matrix $G^{ij} $ was given in \eqref{Gmatrices}.
A correlation function \eqref{corr} was defined in terms of action \eqref{actionsl4}.
The path integral measure is given by
\begin{align}
 \mathcal{D} \Phi = \prod_{i=1}^3 \mathcal{D} \phi_i  \prod_{\alpha =1}^6 \mathcal{D}^2 \beta_\alpha \mathcal{D}^2 \gamma^\alpha  \label{measuresl4}
\end{align}
and the vertex operators are
\begin{align}
 V_\nu (z_\nu) =|\mu^{\nu}_3|^{4(j_3^\nu-j_2^\nu)}|\mu^{\nu}_5|^{4(j_2^\nu-j_1^\nu)}|\mu^{\nu}_6|^{4(j_1^\nu+1)} e^{\mu_\alpha^\nu\gamma^\alpha-\bar{\mu}_\alpha^\nu\bar{\gamma}^\alpha} e^{ \sum_{i=1}^3 2b(j^\nu_i+1)\phi_i } \, . \label{vertexsl4}
\end{align}

In order to reduce the theory to the one corresponding to a free field realization of QSCA, we need to integrate out three sets of ghost systems.
Here we choose to integrate over $\gamma^3$, $\gamma^5$ and $\gamma^6$, since they  do not appear in the interaction terms of \eqref{actionsl4}. This leads to $\delta^{(2)} (\sum_{\nu} \mu_\alpha^\nu )$ and replacements of $\beta_3$, $\beta_5$ and $\beta_6$ by
\begin{equation}\label{f4}
\beta_{\alpha}(z)= - \sum_{\nu=1}^N\frac{\mu^\nu_\alpha}{z-z_\nu}
= - u_\alpha \frac{\prod_{n=1}^{N-2}(z-y_\alpha^n)}{\prod_{\nu=1}^{N}(z-z_\nu)}
\equiv  - u_\alpha \mathcal{B}_\alpha \, ,
\end{equation}
where $\alpha=3,5,6$.
Now the interaction terms become
\begin{equation}\label{d1}
\begin{aligned}
& - \frac 1 {2\pi k}\int d ^2 z\Big[ e ^{b\phi_{1}}\beta_1\bb{1}+ e ^{b\phi_{2}}(\beta_2-\gamma^1\beta_4)(\bb{2}-\gb{1}\bb{4})\\
&\hspace{3cm}-  e ^{b\phi_{3}}(u_3\mathcal{B}_3-\gamma^2u_5\mathcal{B}_5-\gamma^4u_6\mathcal{B}_6)(\bar{u}_3\bar{\mathcal{B}}_3-\bar{\gamma}^2\bar{u}_5\bar{\mathcal{B}}_5-\bar{\gamma}^4\bar{u}_6\bar{\mathcal{B}}_6)\Big] \, .
\end{aligned}
\end{equation}
We make the following change of fields as to remove the $\mathcal{B}_i$-functions
\begin{align}
\begin{aligned}
&\phi_1 + \frac{1}{b} \log |u_5^{-1} u_6 \mathcal{B}_5^{-1} \mathcal{B}_6|^2 \to \phi_1 \, , \\
&\phi_2 + \frac{1}{b} \log |u_3^{-1} u_5 \mathcal{B}_3^{-1} \mathcal{B}_5|^2 \to \phi_2 \, , \\
&\phi_3 + \frac{1}{b} \log |u_3  \mathcal{B}_3 |^2 \to \phi_3 
\end{aligned}
\end{align}
and
\begin{align}
\begin{aligned}
&u_5^{-1} u_6 \mathcal{B}_5^{-1} \mathcal{B}_6 \gamma^1 \to \gamma^1 \, , \quad
 u_5 u_6^{-1} \mathcal{B}_5 \mathcal{B}_6^{-1} \beta_1 \to \beta_1 \, , \\
& u_3^{-1} u_5  \mathcal{B}_3^{-1} \mathcal{B}_5\gamma^2 \to \gamma^2 \, , \quad
u_3 u_5^{-1} \mathcal{B}_3 \mathcal{B}_5^{-1} \beta_2 \to \beta_2 \, , \\
& u_3^{-1} u_6 \mathcal{B}_3^{-1} \mathcal{B}_6 \gamma^4 \to \gamma^4 \, , \quad
u_3 u_6^{-1} \mathcal{B}_3 \mathcal{B}_6^{-1} \beta_4 \to \beta_4 \, .
\end{aligned}
\end{align}
As in subsection \ref{sec:previous}, we formulate $\gamma^\alpha$ and $\beta_\alpha$ as
\begin{align}
\gamma^\alpha (z)\simeq e^{X_\alpha (z)} \eta_\alpha (z)\, , \quad \beta_\alpha (z)= e^{- X_\alpha (z) } \partial \xi_\alpha  (z) \label{ghostb1}
\end{align}
with
\begin{align}
 \eta_\alpha (z) \xi_{\alpha '} (w) \sim \frac{\delta_{\alpha , \alpha '}}{z -w} \, , \quad X_\alpha  (z) X _ {\alpha '} (w) \sim - \delta_{\alpha , \alpha '} \log (z - w) \, .\label{ghostb2}
\end{align}

Analogous to the $\slf(3)$ case, the redefinition of fields provides several contributions to the correlation function.
Taking everything into account, we obtain the following correlator relation
\begin{align} \label{sl4tomin}
 \left \langle \prod^N_{\nu = 1} V_\nu (z_\nu) \right \rangle 
= |\Theta_N|^2 \prod_{\alpha = 3,5,6}\delta ^{(2)} \left( \sum_{\nu=1}^N \mu_\alpha^\nu \right) 
 \left \langle \prod^N_{\nu = 1} \tilde V_\nu (z_\nu) \prod_{n=1}^{N-2} \prod_{\alpha '  =3,5,6} \tilde V^{(\alpha ' )} (y_{\alpha '}^n)  \right \rangle \, ,
\end{align} 
where left-hand side is computed with the action
\begin{align}
\begin{aligned}
& S  = \frac{1}{2 \pi} \int d ^2 z \left[ \frac{G_{ij}}{2} \partial \phi^i \bar \partial \phi^j + \frac{1}{4}\sqrt{g} \mathcal{R} (b (\phi^1 + \phi^2) + (b + b^{-1})\phi^3 ) + \sum_{\alpha =1,2,4} ( \beta_\alpha \bar \partial \gamma^ \alpha+ \bar \beta_\alpha \partial \bar \gamma^\alpha ) \right] \\
&  - \frac 1 {2\pi k}\int d ^2 z\Big[ e ^{b\phi_{1}}\beta_1\bb{1}+ e ^{b\phi_{2}}(\beta_2-\gamma^1\beta_4)(\bb{2}-\gb{1}\bb{4})-  e ^{b\phi_{3}}(1-\gamma^2-\gamma^4)(1-\bar{\gamma}^2-\bar{\gamma}^4)\Big] \, .
 &
\end{aligned} \label{minaction}
\end{align}
The vertex operators at $z_\nu$  receive the shifts of momenta by
\begin{align}
&\tilde V_\nu (z_\nu)  = e^{\sum_{\alpha = 1,2,4} ( \mu_\alpha '{}^{ \nu} \gamma^\alpha -   \bar{\mu}  ' _\alpha {}^  \nu  \bar{\gamma}^\alpha ) } e^{\sum_{i=1}^3 2 b (j_i^\nu + 1) \phi_i  + \phi^3 /b} 
\end{align}
where
\begin{align}
\mu_1 '{}^{ \nu} =  \frac{u_5 \mu_5^\nu  \mu_1^\nu}{ u_6 \mu_6^\nu } \, , \qquad
\mu_2 '{}^{ \nu} =  \frac{u_3 \mu_3^\nu  \mu_2^\nu}{ u_5 \mu_5^\nu } \, , \qquad
\mu_4 '{}^{ \nu} =  \frac{u_3 \mu_3^\nu  \mu_4^\nu}{ u_6 \mu_6^\nu } 
\end{align}
and similarly for $  \bar{\mu}  ' _\alpha$. 
Moreover, there are extra insertions of operators at $y_\alpha^n$ $(\alpha = 3,5,6)$ as
\begin{align}
\begin{aligned}
&\tilde V^{(3)} (y_3^n) = e^{(\phi^2 - \phi^3 ) /b} e^{X_2 + \bar{X}_2 + X_4 + \bar{X}_4  } \, , \\
& \tilde V^{(5)} (y_5^n) = e^{( \phi^1 - \phi^2 ) /b} e^{X_1 + \bar{X}_1 - X_2 - \bar{X}_2 }\, , \\
&\tilde V^{(6)} (y_6^n) = e^{-  \phi^1 /b} e^{ X_1 + \bar{X}_1 + X_4 + \bar{X}_4} \, .
\end{aligned} \label{tvertexsl4}
\end{align}
The prefactor is found to be
\begin{align}
\Theta_N&=u_3^2 \prod_{n > n'}\left[(y_3^n-y_3^{n'})(y_5^n-y_5^{n'})(y_6^n-y_6^{n'})\right]^{2+\frac 3 {4b^2}} \prod_{\mu > \nu}(z_\mu-z_\nu)^{\frac{3}{4b^2}} \\ & \quad \times\prod_{n,m}\left[(y_3^n-y_5^m)(y_3^n-y_6^m)(y_5^n-y_6^m)\right]^{-1-\frac 1 {4b^2}}\prod_{n,\nu}\left[(y_3^n-z_\nu)(y_5^n-z_\nu)(y_6^n-z_\nu)\right]^{-\frac 1 {4b^2}} \, .
\nonumber
\end{align}

It is possible to rewrite the action by changing the basis of ghost systems as
\begin{align}
\begin{aligned}
& \gamma^1 -1 \to \gamma^1 \, , \quad \beta_1 \to \beta_1 \, , \\
&\gamma^2 \to \gamma^2 \, , \quad \beta_2 - \beta_4 \to \beta_2 \, , \\
& \gamma^4 + \gamma^2 - 1 \to \gamma^4 \, , \quad
\beta_4 \to \beta_4 \, . 
\end{aligned}
\end{align}
The kinetic terms of ghost systems are invariant under this change of basis, but the interaction terms are now
\begin{equation}
\begin{aligned}
- \frac 1 {2\pi k}\int d ^2 z\Big[ e ^{b\phi_{1}}\beta_1\bb{1}+ e ^{b\phi_{2}}(\beta_2-\gamma^1\beta_4)(\bb{2}-\gb{1}\bb{4})-  e ^{b\phi_{3}}\gamma^4 \bar \gamma^4\Big] 
\end{aligned}
\end{equation}
from which we observe that the action corresponds to the third realization of QSCA with screening  operators \eqref{screeningmin3}. 

\subsection{Reduction from QSCA to W$^{(2)}_4$-algebra}
\label{sec:QSCAtoW24}

In this subsection, we establish a relation between QSCA  and W$^{(2)}_4$-algebra at the level of correlation functions.  As explained in subsection \ref{sec:QSCA}, there are four types of free field realizations for QSCA. 
Here we shall adopt the first realization with screening charges \eqref{screeningmin1} since it involved the least number of $\gamma$'s.
This free field realization of QSCA can be described by a theory with the action
\begin{align}\label{f2}
\begin{aligned}
 S &= \frac 1 {2\pi}\int d^2 z\Big[\frac{G^{ij}}{2}\p\phi_{i}\pb\phi_{j}+\frac{1}{4}\sqrt{g}\mathcal{R}(b (\phi^1+\phi^2) + (b + b^{-1})\phi^3)+\sum_{\alpha=1}^3(\beta_\alpha \pb\gamma^\alpha+\bb{\alpha} \partial\gb{\alpha})\Big]\\
& \quad - \frac 1 {2\pi k}\int d^2 z\Big[e^{b\phi_{1}}\beta_1\bb{1}+e^{b\phi_{2}}(\beta_2-\gamma^1\beta_3)(\bb{2}-\gb{1}\bb{3})- e^{b\phi_{3}}\Big] \, .
\end{aligned}
\end{align}
Consider a correlation function of the form \eqref{corr}, with the action \eqref{f2} and
the path integral measure
\begin{align}
 \mathcal{D} \Phi= \prod_{i=1}^3 \mathcal{D} \phi_i  \prod_{\alpha =1}^3 \mathcal{D}^2 \beta_\alpha \mathcal{D}^2 \gamma^\alpha.  \label{measuremin}
\end{align}
The vertex operators take the form
\begin{align}
 V_\nu (z_\nu) = |\mu^{\nu}_2|^{4(j_2^\nu + 1)} e^{\mu_\alpha^\nu\gamma^\alpha-\bar{\mu}_\alpha^\nu\bar{\gamma}^\alpha} e^{ \sum_{i=1}^3 2b(j^\nu_i+1)\phi_i } \, . \label{vertexmin}
\end{align}

Now we want to reduce the theory to the one corresponding to a free field realization of W$^{(2)}_4$-algebra by integrating out two sets of ghost systems.
Since $\gamma^1$ appears in the interaction terms of the action \eqref{f2}, we integrate with respect to $\gamma^2$ and $\gamma^3$. This yields the delta functions $\delta^{(2)} (\sum_{\nu} \mu_2^\nu) \delta^{(2)} (\sum_{\nu} \mu_3^\nu)$ and the relations
\begin{equation}\label{a7}
\beta_{\alpha}(z)= - \sum_{\nu=1}^N\frac{\mu^\nu_\alpha}{z-z_\nu}= - u_\alpha \frac{\prod_{n=1}^{N-2}(z-y_\alpha^n)}{\prod_{\nu=1}^{N}(z-z_\nu)} \equiv - u_\alpha \mathcal{B}_\alpha 
\end{equation}
where $\alpha =2,3$.
Shifting $\phi_2 $ by
\begin{equation}\label{a9}
\phi_2(z) + \frac 1 b \ln{|u_2 \mathcal{B}_2|^2} \to \phi_2 (z) \, , 
\end{equation}
the interaction terms become
\begin{align}
- \frac 1 {2\pi k}\int \dd^2 z\Big[e^{b\phi_{1}} \beta_1 \bar \beta_1- e^{b\phi_{2}} (1-u_2^{-1} u_3 \mathcal{B}_2^{-1}\mathcal{B}_3\gamma^1 ) (1-\bar u_2^{-1} \bar u_3 \bar {\mathcal{B}}_2^{-1}\bar{\mathcal{B}}_3 \bar \gamma^1 )  - e^{b\phi_{3}}\Big] \, .
\end{align}
We further rescale $\gamma^1$ and $\beta_1$ by 
\begin{equation}
u_2^{-1} u_3\mathcal{B}_2^{-1}\mathcal{B}_3 \gamma^1 \to \gamma^1 \, , \quad
u_2 u_3^{-1} \mathcal{B}_2 \mathcal{B}_3^{-1} \beta_1 \to \beta_1 
\end{equation}
and reformulate the ghosts in terms of a pair of free fermions and a free boson as we did in \eqref{ghostb1} and \eqref{ghostb2}.

We then arrive at the correlator relation as
\begin{align} \label{mintosub}
\left \langle \prod^N_{\nu = 1} V_\nu (z_\nu) \right \rangle   = |\Theta_N|^2 \delta ^{(2)} \left( \sum_{\nu=1}^N \mu_2^\nu \right) \delta ^{(2)} \left( \sum_{\nu=1}^N \mu_3^\nu \right) 
 \left \langle \prod^N_{\nu = 1} \tilde V_\nu (z_\nu) \prod_{n=1}^{N-2} {\tilde V}^{(2)} (y_2^n)  {\tilde V}^{(3)} (y_3^n) \right \rangle \, ,
\end{align} 
whose right-hand side is evaluated with the action
\begin{align}
\begin{aligned}
& S   = \frac{1}{2 \pi} \int d ^2 z \left[ \frac{G_{ij}}{2} \partial \phi^i \bar \partial \phi^j + \frac{1}{4}\sqrt{g} \mathcal{R} ( b \phi^1 + (b + b^{-1})( \phi^2 + \phi^3 )) +  \beta_1 \bar \partial \gamma^1 + \bar \beta_1 \partial \bar \gamma^1 ) \right] \\
& \qquad   - \frac 1 {2\pi k}\int d ^2 z\Big[ e ^{b\phi_{1}}\beta_1\bb{1} -  e ^{b\phi_{2}}(1-\gamma^1 )(1 - \bar { \gamma }^1)-  e ^{b\phi_{3}}\Big] \, .
 &
\end{aligned} \label{subaction}
\end{align}
If we make the shift  $\gamma^1 -1 \to \gamma^1$ in \eqref{subaction}, the action coincides with that of the first realization of W$^{(2)}_4$-algebra whose screening operators are given in \eqref{screeningW241}.
The new vertex operators in the correlation function \eqref{mintosub} are found to be
\begin{align}
\begin{aligned}
&\tilde V_\nu (z_\nu)  = e^{ \mu_1 '{}^{ \nu} \gamma^1 -   \bar{\mu}  ' _1 {}^  \nu  \bar{\gamma}^1} e^{\sum_{i=1}^3 2 b (j_i^\nu + 1) \phi_i + \phi^2 /b} \, ,  \\
& \tilde V^{(2)} (y_2^n) = e^{ - \phi^2  /b}  e^{ X_1 + \bar X_1 } \, , \quad
 \tilde V^{(3)} (y_3^n) =  e^{-X_1- \bar X_1} 
\end{aligned} \label{tvertexsub}
\end{align}
where
\begin{align}
\mu_1 '{}^{ \nu} =  \frac{u_2 \mu_2^\nu  \mu_1^\nu}{ u_3 \mu_3^\nu } \, , \quad
\bar{\mu}_1 '{}^{ \nu} =  \frac{\bar{u}_2 \bar{\mu}_2^\nu  \bar{\mu}_1^\nu}{ \bar{u}_3 \bar{\mu}_3^\nu } \, .
\end{align}
The prefactor is
\begin{align}
\Theta_N&=u_2^3 u_3^{-1} \prod_{n > n'}(y_2^n-y_2^{n'}) ^{1+\frac{1} {b^2}}  (y_3^n-y_3^{n'})\prod_{\mu > \nu}(z_\mu-z_\nu)^{\frac{1}{b^2}} \prod_{n,m}(y_2^n-y_3^m)^{-1}\prod_{n,\nu}(y_2^n-z_\nu)^{-\frac 1 {b^2}} \, .
\end{align}

\subsection{Reduction from W$^{(2)}_4$-algebra  to W$_4$-algebra}

In this subsection, we study the reduction relation  from  W$^{(2)}_4$-algebra to W$_4$-algebra. As discussed in subsection \ref{sec:W24}, there are two free field realizations for W$^{(2)}_4$-algebra. For the purpose of this subsection, we use the second realization with screening charges \eqref{screeningW242}.
The correlation function takes usual form \eqref{corr} with the action
\begin{align}
\begin{aligned}
 S &= \frac 1 {2\pi}\int d^2 z\Big[\frac{G^{ij}}{2}\p\phi_{i}\pb\phi_{j}+\frac{1}{4}\sqrt{g}\mathcal{R}( b \phi^1+ (b + b^{-1})(\phi^2 + \phi^3) )+ \beta  \pb\gamma +\bar \beta \partial \bar \gamma \Big]\\
& \quad - \frac 1 {2\pi k}\int d^2 z\Big[e^{b\phi_{1}}\beta \bar \beta - e^{b\phi_{2}} - e^{b\phi_{3}}\Big]  \, 
\end{aligned}
\end{align}
and the path integral measure.
\begin{align}
 \mathcal{D} g = \prod_{i=1}^3 \mathcal{D} \phi_i   \mathcal{D}^2 \beta \mathcal{D}^2 \gamma \, . \label{measuresub}
\end{align}
The vertex operators are given by
\begin{align}
 V_\nu (z_\nu) = |\mu^{\nu}|^{4(j_1^\nu + 1)} e^{\mu^\nu\gamma -\bar{\mu}^\nu\bar{\gamma}} e^{ \sum_{i=1}^3 2b(j^\nu_i+1)\phi_i } \, . \label{vertexsub}
\end{align}

The reduction procedure follows similarly as in previous cases. We integrate out $\gamma$ and $\beta$ from the correlation function for W$^{(2)}_4$-algebra, which yields
a delta function $\delta^{(2)} (\sum_\nu \mu^\nu )$ and the following expression of $\beta(z)$ in terms of function $\mathcal{B}$: 
\begin{equation}
\beta (z)= - \sum_{\nu=1}^N\frac{\mu^\nu}{z-z_\nu}= - u \frac{\prod_{n=1}^{N-2}(z-y^n)}{\prod_{\nu=1}^{N}(z-z_\nu)} \equiv - u \mathcal{B}(z,y^n,z_\nu) \, .
\end{equation}
We further shift  $\phi_1(z)$ by
\begin{equation}
\phi_1(z) + \frac 1 b\ln{|u \mathcal{B}|^2} \to \phi_1 (z)\, .
\end{equation}
The final form of the correlator is given by
\begin{align} \label{subtoW4}
\begin{aligned}
& \left \langle \prod^N_{\nu = 1} V_\nu (z_\nu) \right \rangle   = |\Theta_N|^2 \delta ^{(2)} \left( \sum_{\nu=1}^N \mu^\nu \right) 
 \left \langle \prod^N_{\nu = 1} \tilde V_\nu (z_\nu) \prod_{n=1}^{N-2} {\tilde V}_b (y^n)  \right \rangle \, ,
\end{aligned}
\end{align} 
where the action for the right-hand side is
\begin{align}
& S  = \frac{1}{2 \pi} \int d ^2 z \left[ \frac{G_{ij}}{2} \partial \phi^i \bar \partial \phi^j + \frac{Q_\phi}{4}\sqrt{q} \mathcal{R} (\phi^1 + \phi^2 + \phi^3 )  + \frac { 1 }{k} \left( e ^{b\phi_{1}} + e ^{b\phi_{2}} + e ^{b\phi_{3}} \right) \right] 
\end{align}
with $Q_\phi = b + b^{-1}$. This is nothing but the action of $\slf(4)$ Toda field theory.
The new vertex operators are
\begin{equation}
\tilde V_\nu (z_\nu)  =  e^{\sum_{i=1}^3 2 b (j_i^\nu + 1) \phi_i  + \phi^1 /b} \, ,  \qquad \tilde V _b (y^n) = e^{ - \phi^1  /b}   \, .
 \label{tvertexw4}
\end{equation}
And the prefactor is found to be
\begin{align}
\Theta_N&=u^2 \prod_{n > n'}(y^n-y^{n'}) ^{\frac{3} {4 b^2}}  \prod_{\mu > \nu}(z_\mu-z_\nu)^{\frac{3}{4 b^2}} \prod_{n,\nu}(y^n-z_\nu)^{-\frac{3}{4 b^2}} \, .
\end{align}

\section{Conclusion and discussions}
\label{sec:conclusion}

In this paper, we derived new correspondences among correlation functions of theories with W-algebra symmetry.
We generalize the previous works in \cite{Ribault:2005wp,Hikida:2007tq,Hikida:2007sz,Creutzig:2011qm,Creutzig:2015hla}, where $\slf(N)$ WZNW model is reduced to a theory with QSCA symmetry. The screening charges constructed in  \cite{Genra1,Genra2, CGN} are employed to develop free field realizations of non-regular W-algebras.

The paper started with a detailed account of W-algebras associated with $\slf(3)$.
The non-regular W-algebra in this case, BP-algebra, has two different free field realizations. The realization with the screening operators \eqref{screening3} was reduced to $\slf(3)$ Toda theory as in \eqref{BPtoW3} using path integral. 
A new method of putting restrictions on momenta of vertex operators was proposed in order to obtain correlation relations such as in \eqref{sl3toBP2}. The analysis was then extended to the study of W-algebras associated to $\slf(4)$. In this case, there are three types of non-regular W-algebras and more complicated correlator relations were derived. Similarly to the $\slf(3)$ case, we started from the construction of free field realizations for the non-regular W-algebras using the screening charges in appendix \ref{sec:screening}. The method developed for the $\slf(3)$ case were applied directly in deriving the correlator relations for the  $\slf(4)$ case. In particular, we derived new correlator correspondences in several explicit examples such as in \eqref{sl4tomin}, \eqref{mintosub}, and \eqref{subtoW4}. The focus of the paper is to examine such examples of W-algebras associated to $\slf(3)$ and $\slf(4)$. An attempt of generalize the results to the $\slf(N)$ case was made, with correlator relations for a few special cases presented in appendix  \ref{sec:corrslN}. Further generalizations of such correlator relations to W-algebras associated with any Lie algebra $\g$ remain open. The screening charges for free field realizations of W-algebras associated with $\mathfrak{so}(5)$, for example, are given in appendix \ref{sec:screening}. It is straightforward to apply the techniques developed in this paper and derive the correlator relations for the $\mathfrak{so}(5)$ case. 


Correlator relations like the ones derived in this paper have a wide range of applications.
We aim to report on them in the near future.
A main application of the original correlator relation with $\slf(2)$ is the proof of the Fateev-Zamolodchikov-Zamolodchikov (FZZ) duality conjecture in \cite{Hikida:2008pe}. See also \cite{Creutzig:2010bt}.
We expect that generalized FZZ dualities can be derived from our new correlator relations presented in this paper.
As another application, recall that structure constants of the operator algebra for $\mathfrak{osp}(1|2)$ WZNW model are determined from those of $\mathcal{N}=1$ super Liouville field theory \cite{Hikida:2007sz, Creutzig:2010zp}. Given that the structure constants for $\slf(N)$ Toda field theory have been computed in \cite{Fateev:2007ab,Fateev:2008bm}, and the correlator relation between W$^{(2)}_N$-algebra and W$_N$-algebra has been established  in this paper, it is possible to apply an analogous procedure as in \cite{Hikida:2007sz, Creutzig:2010zp} and obtain the structure constants for W$^{(2)}_N$-algebra.

As mentioned in the introduction, non-regular W-algebras have received a lot of attention, though there is much that remains poorly understood.
For example, non-regular W-algebras arise if surface operators  are inserted in four dimensional $SU(N)$ gauge theories.
It was suggested that correlator correspondences like those presented in this paper can be obtained from different treatments of the same surface operators \cite{Alday:2010vg,Kozcaz:2010yp,Wyllard:2010rp,Wyllard:2010vi}. It is important to compare and establish direct relations between correlator correspondences obtained from four dimensional gauge theory with those from our path integral derivation.
As another example, non-regular W-algebras can be realized as an asymptotic symmetry of higher spin gravity with non-standard gravitational sector. The understanding of non-regular W-algebra from W$_N$-algebra is expected to help with examining higher spin gravity. In particular, we would like to study the properties of conical defect geometry in higher spin gravity as in \cite{Castro:2011iw,Gaberdiel:2012ku,Perlmutter:2012ds,Hikida:2012eu}.
A partial result for this has been already provided in \cite{Creutzig:2019qos}.
It is also of our great interest to introduce supersymmetry to relate superstring theory as in \cite{Creutzig:2011fe,Creutzig:2013tja,Creutzig:2014ula,Eberhardt:2018plx,Creutzig:2019qos}, see also \cite{Gaberdiel:2013vva,Gaberdiel:2014cha}.

\subsection*{Acknowledgements}

We are grateful to David Ridout and Zac Fehily for useful discussions. We thank the organiser of workshop 'Vertex Operator Algebras and Related Topics in Kumamoto', where a part of the work was done. 
The work of TC is supported by NSERC grant number RES0019997.
The work of NG is supported by JSPS Overseas Research Fellowships.
The work of YH is supported by JSPS KAKENHI Grant Number 16H02182 and 19H01896.
The work of TL is supported by JSPS KAKENHI Grant Number 16H02182. TL would like to thank the support of AustMS Lift-Off Fellowship to visit collaborators.

\appendix

\section{Screening operators of $\W^k(\g,f)$}
\label{sec:screening}

Let $\g$ be a simple Lie algebra of $\dim\g<\infty$ with the bilinear form $(\cdot|\cdot)$ normalized such that the longest root has norm two, $\h$ a Cartan subalgebra of $\g$ and $\widehat{\g}$ the affine Lie algebra of $\g$. Denote by $\Delta$, $\Delta_+$, $\Pi=\{\alpha_1,\ldots,\alpha_l\}$ sets of roots, positive roots and simple roots respectively of $\g$, where $l=\dim\h$. Let $M_{\g}$ be the Weyl vertex algebra associated to $\g$, which is generated by fields $\beta_\alpha(z)$, $\gamma_\alpha(z)$ for all $\alpha\in\Delta_+$ satisfying that
\begin{align}
\beta_{\alpha}(z)\gamma_{\alpha'}(w)\sim - \frac{\delta_{\alpha,\alpha'}}{z-w} \, ,\quad
\beta_{\alpha}(z)\beta_{\alpha'}(w)\sim0\sim\gamma_{\alpha}(z)\gamma_{\alpha'}(w) \, ,\quad
\alpha,\alpha'\in\Delta_+ \, .
\end{align}
Then $M_\g$ is isomorphic to $|\Delta_+|$ tensor products of $\beta\gamma$-systems. Let $\pi$ be the Heisenberg vertex algebra associated with $\h$, which is generated by fields $\alpha_i(z)$ for all $i=1,\ldots,l$ satisfying that
\begin{align}
\alpha_i(z)\alpha_j(w)\sim\frac{(\alpha_i|\alpha_j)}{(z-w)^2} \, ,\quad
i, j=1,\ldots,l \, .
\end{align}
Let $V^k(\g)$ be the (universal) affine vertex algebra of $\g$ at level $k\in\C$. By Wakimoto (for $\g=\slf(2)$), Feigin and Frenkel \cite{Wakimoto, FF88}, we have the following free field realizations, called Wakimoto representations of $\widehat{\g}$:
\begin{align}
V^k(\g)\hookrightarrow M_{\g}\otimes\pi \, ,
\end{align}
whose image coincides with the common kernel of screening operators $S_i$ for $i=1,\ldots,l$ if $k$ is generic \cite{Frenkel}:
\begin{align}
V^k(\g)\simeq\bigcap_{i=1}^l\Ker S_i\subset M_{\g}\otimes\pi\, .
\end{align}

For any nilpotent element $f$ of $\g$, one can define the (affine) W-algebra $\W^k(\g,f)$ as follows. Let $x_0$ be a semisimple element of $\g$ and
\begin{align}
\Gamma:\g=\bigoplus_{j\in\frac{1}{2}\Z}\g_j \, ,\quad
\g_j=\{u\in\g\mid[x_0,u]=j u\}
\end{align}
a $\frac{1}{2}\Z$-grading of $\g$ such that $f\in\g_{-1}$ and $\ad(f)\colon\g_j\rightarrow\g_{j-1}$ is injective for $j\geq\frac{1}{2}$ and surjective for $j\leq\frac{1}{2}$. Then $\Gamma$ is called a good grading for $f$. By Jacobson-Morozov theorem, there exists a $\slf(2)$-triple $\{e_0,h_0,f\}$ in $\g$ and $x_0=\frac{1}{2}h_0$ defines a good grading of $\g$ for $f$. Hence good gradings exist for any $f$. These good gradings are classified by weighted Dynkin diagrams \cite{EK}, which are Dynkin diagram with weights of $\alpha_i$ in $\{0,\frac{1}{2},1\}$ for each $\alpha_i\in\Pi$. Kac, Roan and Wakimoto \cite{KRW} define the $\W$-algebras $\W^k(\g,f)$ as the (generalized) Drinfeld-Sokolov reduction of $V^k(\g)$ associated to $(\g,f,\Gamma)$, which is a generalization of results by Feigin and Frenkel \cite{FF90}:
\begin{align}
\W^k(\g,f)=H^0_{DS,f}(V^k(\g))\, .
\end{align}
Then $\W^k(\g,f)$ is a $\frac{1}{2}\Z_{\geq0}$-graded vertex algebra. This construction depends on the choice of $\Gamma$, but the vertex algebras obtained by these procedures are isomorphic to each other for any choice of $\Gamma$ with fixed $\g$ and $f$. The difference of choice of $\Gamma$ only appears in conformal degrees on $\W^k(\g,f)$, or equivalently, the choice of Virasoro field in $\W^k(\g,f)$.

\smallskip

Applying the functor $H^0_{DS,f}(?)$ to the embedding of $V^k(\g)$ into $M_{\g}\otimes\pi$, we have free field realizations of $\W^k(\g,f)$ \cite{Genra2}:
\begin{align}
\W^k(\g,f)\hookrightarrow\widetilde{M}_\g\otimes\pi \, ,
\end{align}
where $\widetilde{M}_\g$ is a vertex algebra isomorphic to the $(\dim\g_0+\frac{1}{2}\dim\g_{\frac{1}{2}})$ tensor products of the $\beta\gamma$ system. If $k$ is generic, the image coincides with the common kernel of screening operators $Q_i$, which is an intertwining operator induced by $S_i$, for $i=1,\ldots,l$:
\begin{align}
\W^k(\g,f)\simeq\bigcap_{i=1}^l\Ker Q_i\subset\widetilde{M}_\g\otimes\pi \, .
\end{align}

\subsection{Screening operators of $\W^k(\slf(4),f)$}
Consider the case $\g=\slf(4)$. Then
\begin{align}
\Pi=\{\alpha_1,\alpha_2,\alpha_3\},\quad
\Delta_+=\{\alpha_1,\alpha_2,\alpha_3,\alpha_1+\alpha_2,\alpha_2+\alpha_3,\alpha_1+\alpha_2+\alpha_3\}\, .
\end{align}
Hence $M_{\slf(4)}$ is isomorphic to $6$ tensor products of the $\beta\gamma$-system. Set
\begin{align*}
&\beta_1=\beta_{\alpha_1} \, ,\quad
\beta_2=\beta_{\alpha_2} \, ,\quad
\beta_3=\beta_{\alpha_3} \, ,\quad
\beta_4=\beta_{\alpha_1+\alpha_2} \, ,\quad
\beta_5=\beta_{\alpha_2+\alpha_3}\, ,\quad
\beta_6=\beta_{\alpha_1+\alpha_2+\alpha_3} \, ,\\
&\gamma_1=\gamma_{\alpha_1}\, ,\quad
\gamma_2=\gamma_{\alpha_2}\, ,\quad
\gamma_3=\gamma_{\alpha_3}\, ,\quad
\gamma_4=\gamma_{\alpha_1+\alpha_2}\, ,\quad
\gamma_5=\gamma_{\alpha_2+\alpha_3}\, ,\quad
\gamma_6=\gamma_{\alpha_1+\alpha_2+\alpha_3} \, .
\end{align*}
Here and further on, we often drop the indeterminate $z$, e.g., we shall write $\beta$ in place of $\beta(z)$. Wakimoto free field realizations of $V^k(\slf (4))$ are the following embeddings
\begin{align}
V^k(\slf(4))\hookrightarrow M_{\slf(4)}\otimes\pi\, ,
\end{align}
where
\begin{align}
M_{\slf(4)}&=\langle\beta_i,\gamma_i\mid i=1,\ldots,6 \rangle_{\mathrm{VA}}\, ,\quad
\beta_i(z)\gamma_j(w)\sim\frac{\delta_{i,j}}{z-w}\, ,\\
\pi&=\langle\alpha_i\mid i=1,2,3 \rangle_{\mathrm{VA}\, },\quad
\alpha_i(z)\alpha_j(w)\sim\frac{(\alpha_i|\alpha_j)}{(z-w)^2}\, .
\end{align}
If $k$ is generic, the image may be described as the common kernel of $3$ screening operators $S_1$, $S_2$, $S_3$. Given a nilpotent element $f$ of $\slf(4)$ and a good grading on $\slf(4)$ for $f$, these $S_i$ induce screening operators $Q_1$, $Q_2$, $Q_3$ of $\W^k(\slf(4),f)$ via the Drinfeld-Sokolov reduction. Indeed, we have several choice of $S_i$, see below. If we choose suitable $S_i$ for each $i$ with respect to $f$, we can compute $Q_i$ explicitly.

\smallskip

Using these reduction techniques, we give examples of screening operators $Q_i$ of $\W^k(\slf(4),f)$. In $\slf(4)$, there exist $5$ nilpotent orbits, classified by partitions of $4$:
\begin{align}
(4)\, ,\quad
(3,1 )\, ,\quad
(2^2)\, ,\quad
(2,1^2 )\, ,\quad
(1^4)\, .
\end{align}
Then principal, subregular, minimal, zero nilpotent orbits correspond to $(4)$, $(3,1)$, $(2,1^2)$, $(1^4)$ respectively. In case $f=0$, the $\W^k(\slf(4),f)$ is just the affine vertex algebra $V^k(\slf(4))$. Hence we only need to consider the case $f$ belongs to one of the following nilpotent orbits: principal, subregular, $(2^2)$ and minimal. We will use the following notation
\begin{align}
\kappa=\sqrt{k+4}\, .
\end{align}

\subsubsection*{Principal nilpotent}
Consider a principal nilpotent element
\begin{align}
f=f_{\alpha_1}+f_{\alpha_2}+f_{\alpha_3}
\end{align}
with the (unique) good grading
\begin{align}
\setlength{\unitlength}{1mm}
\begin{picture}(0,0)(20,10)
\put(10,10){\circle{2}}
\put(9,13){\footnotesize$1$}
\put(9,5){\footnotesize$\alpha_1$}
\put(11,10.3){\line(1,0){8}}
\put(20,10){\circle{2}}
\put(19,13){\footnotesize$1$}
\put(19,5){\footnotesize$\alpha_2$}
\put(21,10.3){\line(1,0){8}}
\put(30,10){\circle{2}}
\put(29,13){\footnotesize$1$}
\put(29,5){\footnotesize$\alpha_3$}
\put(32.5,9){.}
\end{picture}
\end{align}
If $k$ is generic,
\begin{align}
V^k(\slf (4) )\simeq\Ker S_1\cap\Ker S_2\cap\Ker S_3\subset M_{\slf (4) }\otimes\pi\, ,
\end{align}
where
\begin{align}\nonumber
S_1&=\int \left(\beta_1+\frac{1}{2}\gamma_2\beta_4+\left(\frac{1}{2}\gamma_5+\frac{1}{12}\gamma_2\gamma_3\right)\beta_6\right)\ \e^{-\kappa^{-1}\int \alpha_1(z)} dz\, ,\\
S_2&=\int \left(\beta_2-\frac{1}{2}\gamma_1\beta_4+\frac{1}{2}\gamma_3\beta_5-\frac{1}{6}\gamma_1\gamma_3\beta_6\right)\ \e^{-\kappa^{-1}\int \alpha_2(z)} dz\, ,\\
S_3&=\int \left(\beta_3-\frac{1}{2}\gamma_2\beta_5+\left(-\frac{1}{2}\gamma_4+\frac{1}{12}\gamma_1\gamma_2\right)\beta_6\right)\ \e^{-\kappa^{-1}\int \alpha_3(z)} dz\, .\nonumber
\end{align}
As before we assume that every expression is normally ordered and we omit the normally ordered product signs. 
Applying the Drinfeld-Sokolov reduction, we have
\begin{align}
\W^k(\slf (4) )=\W^k(\slf (4) ,\mathrm{principal})\simeq\Ker Q_1\cap\Ker Q_2\cap\Ker Q_3\subset \pi\, ,
\end{align}
where
\begin{align} \nonumber
Q_1&=\int \e^{-\kappa^{-1}\int \alpha_1(z)}dz\, ,\\
Q_2&=\int \e^{-\kappa^{-1}\int \alpha_2(z)} dz\, ,\\
Q_3&=\int \e^{-\kappa^{-1}\int \alpha_3(z)} dz\, . \nonumber
\end{align}

\subsubsection*{Subregular nilpotent}
Consider subregular nilpotent elements
\begin{align}
f_1=f_{\alpha_2}+f_{\alpha_3},\quad
f_2=f_{\alpha_1+\alpha_2}+f_{\alpha_3}
\end{align}
with the good grading
\begin{align}
\setlength{\unitlength}{1mm}
\begin{picture}(0,0)(20,10)
\put(10,10){\circle{2}}
\put(9,13){\footnotesize$0$}
\put(9,5){\footnotesize$\alpha_1$}
\put(11,10.3){\line(1,0){8}}
\put(20,10){\circle{2}}
\put(19,13){\footnotesize$1$}
\put(19,5){\footnotesize$\alpha_2$}
\put(21,10.3){\line(1,0){8}}
\put(30,10){\circle{2}}
\put(29,13){\footnotesize$1$}
\put(29,5){\footnotesize$\alpha_3$}
\put(32.5,9){.}
\end{picture}
\end{align}
If $k$ is generic,
\begin{align}
V^k(\slf(4))\simeq\Ker S_1\cap\Ker S_2\cap\Ker S_3\subset M_{\slf(4)}\otimes\pi\, ,
\end{align}
where
\begin{align} \nonumber
S_1&=\int \beta_1\ \e^{-\kappa^{-1}\int \alpha_1(z)} dz\, ,\\
S_2&=\int \left(\beta_2-\gamma_1\beta_4+\gamma_3\beta_5\right)\ \e^{-\kappa^{-1}\int \alpha_2(z)} dz\, ,\\
S_3&=\int \left(\beta_3-\gamma_4\beta_6\right)\ \e^{-\kappa^{-1}\int \alpha_3(z)} dz\, . \nonumber
\end{align}
Applying the Drinfeld-Sokolov reductions associated to $f=f_1,f_2$, we have
\begin{align}
\W^k(\slf (4) ,\mathrm{subregular})\simeq\Ker Q_1\cap\Ker Q_2\cap\Ker Q_3\subset \langle\beta_1\, ,\gamma_1\rangle_{\mathrm{VA}}\otimes\pi,
\end{align}
where
\begin{align} \label{subs}
\begin{aligned}
Q_1&=\int \beta_1\ \e^{-\kappa^{-1}\int \alpha_1(z)} dz\, ,\\
Q_2&=
\begin{cases}
\displaystyle \int \e^{-\kappa^{-1}\int \alpha_2(z)} dz & \mathrm{if}\ f=f_1\, ,\\
\displaystyle \int \gamma_1\ \e^{-\kappa^{-1}\int \alpha_2(z)} dz & \mathrm{if}\ f=f_2\, ,
\end{cases}
\\
Q_3&=\int \e^{-\kappa^{-1}\int \alpha_3(z)} dz\, . 
\end{aligned}
\end{align}

\subsubsection*{Rectangular nilpotent of type ($2^2$)}
Consider rectangular nilpotent elements
\begin{align}
f_1=f_{\alpha_1+\alpha_2}+f_{\alpha_2+\alpha_3}\, ,\quad
f_2=f_{\alpha_2}+f_{\alpha_1+\alpha_2+\alpha_3}
\end{align}
of Jordan type $(2^2)$ with the good grading
\begin{align}
\setlength{\unitlength}{1mm}
\begin{picture}(0,0)(20,10)
\put(10,10){\circle{2}}
\put(9,13){\footnotesize$0$}
\put(9,5){\footnotesize$\alpha_1$}
\put(11,10.3){\line(1,0){8}}
\put(20,10){\circle{2}}
\put(19,13){\footnotesize$1$}
\put(19,5){\footnotesize$\alpha_2$}
\put(21,10.3){\line(1,0){8}}
\put(30,10){\circle{2}}
\put(29,13){\footnotesize$0$}
\put(29,5){\footnotesize$\alpha_3$}
\put(32.5,9){.}
\end{picture}
\end{align}
If $k$ is generic,
\begin{align}
V^k(\slf (4) )\simeq\Ker S_1\cap\Ker S_2\cap\Ker S_3\subset M_{\slf (4) }\otimes\pi\, ,
\end{align}
where
\begin{equation}
\begin{aligned}\label{i3}
S_1&=\int \beta_1\ \e^{-\kappa^{-1}\int \alpha_1(z)} dz\, ,\\
S_2&=\int \left(\beta_2-\gamma_1\beta_4+\gamma_3\beta_5-\gamma_1\gamma_3\beta_6\right)\ \e^{-\kappa^{-1}\int \alpha_2(z)} dz\, ,\\
S_3&=\int \beta_3\ \e^{-\kappa^{-1}\int \alpha_3(z)} dz\, .
\end{aligned}
\end{equation}
Applying the Drinfeld-Sokolov reductions associated to $f=f_1,f_2$, we have
\begin{align}
\W^k(\slf (4) ,(2^2))\simeq\Ker Q_1\cap\Ker Q_2\cap\Ker Q_3\subset \langle\beta_1,\gamma_1,\beta_3,\gamma_3\rangle_{\mathrm{VA}}\otimes\pi \, ,
\end{align}
where
\begin{equation}
\begin{aligned}\label{i2}
Q_1&=\int \beta_1\ \e^{-\kappa^{-1}\int \alpha_1(z)}:dz\, ,\\
Q_2&=
\begin{cases}
\displaystyle \int \left(\gamma_1-\gamma_3\right)\ \e^{-\kappa^{-1}\int \alpha_2(z)} dz & \mathrm{if}\ f=f_1\, ,\\
\displaystyle \int \left(1-\gamma_1\gamma_3\right)\ \e^{-\kappa^{-1}\int \alpha_2(z)} dz & \mathrm{if}\ f=f_2\, ,
\end{cases}
\\
Q_3&=\int \beta_3\ \e^{-\kappa^{-1}\int \alpha_3(z)} dz\, .
\end{aligned}
\end{equation}

\subsubsection*{Minimal nilpotent}

Consider minimal nilpotent elements
\begin{align}
f_1=f_{\alpha_3}\, ,\quad
f_2=f_{\alpha_2+\alpha_3}\, ,\quad
f_3=f_{\alpha_1+\alpha_2+\alpha_3}\, .
\end{align}

\bigskip

\noindent {\bf $\Z$-graded case.}
We have a good grading for any $f_i$:
\begin{align}
\setlength{\unitlength}{1mm}
\begin{picture}(0,0)(20,10)
\put(10,10){\circle{2}}
\put(9,13){\footnotesize$0$}
\put(9,5){\footnotesize$\alpha_1$}
\put(11,10.3){\line(1,0){8}}
\put(20,10){\circle{2}}
\put(19,13){\footnotesize$0$}
\put(19,5){\footnotesize$\alpha_2$}
\put(21,10.3){\line(1,0){8}}
\put(30,10){\circle{2}}
\put(29,13){\footnotesize$1$}
\put(29,5){\footnotesize$\alpha_3$}
\put(32.5,9){.}
\end{picture}
\end{align}
If $k$ is generic,
\begin{align}
V^k(\slf(4))
\simeq\Ker S_1\cap\Ker S_2\cap\Ker S_3
\simeq\Ker \widetilde{S}_1\cap\Ker \widetilde{S}_2\cap\Ker \widetilde{S}_3\subset M_{\slf(4)}\otimes\pi\, ,
\end{align}
where
\begin{align} \label{sl4min}
\begin{aligned}
S_1&=\int \beta_1\ \e^{-\kappa^{-1}\int \alpha_1(z)} dz\, ,\\
S_2&=\int \left(\beta_2-\gamma_1\beta_4\right)\ \e^{-\kappa^{-1}\int \alpha_2(z)} dz\, ,\\
S_3&=\int \left(\beta_3-\gamma_2\beta_5-\gamma_4\beta_6\right)\ \e^{-\kappa^{-1}\int \alpha_3(z)} dz
\end{aligned}
\end{align}
and
\begin{align} \nonumber
\widetilde{S}_1&=\int \left(\beta_1+\frac{1}{2}\gamma_2\beta_4\right)\ \e^{-\kappa^{-1}\int \alpha_1(z)} dz\, ,\\
\widetilde{S}_2&=\int \left(\beta_2-\frac{1}{2}\gamma_1\beta_4\right)\ \e^{-\kappa^{-1}\int \alpha_2(z)} dz\, ,\\
\widetilde{S}_3&=\int \left(\beta_3-\gamma_2\beta_5+\left(\frac{1}{2}\gamma_1\gamma_2-\gamma_4\right)\beta_6\right)\ \e^{-\kappa^{-1}\int \alpha_3(z)} dz \, . \nonumber
\end{align}
Applying the Drinfeld-Sokolov reductions associated to $f=f_1,f_2,f_3$, we have
\begin{equation}
\begin{split}
\W^k(\slf (4) ,\mathrm{minimal})
&\simeq\Ker Q_1\cap\Ker Q_2\cap\Ker Q_3
\simeq\Ker \widetilde{Q}_1\cap\Ker \widetilde{Q}_2\cap\Ker \widetilde{Q}_3\\
&\subset \langle\beta_1,\gamma_1,\beta_2,\gamma_2,\beta_4,\gamma_4\rangle_{\mathrm{VA}}\otimes\pi\, ,
\end{split}
\end{equation}
where
\begin{align} \label{QSCAi}
\begin{aligned}
Q_1&=\int \beta_1\ \e^{-\kappa^{-1}\int \alpha_1(z)} dz\, ,\\
Q_2&=\int \left(\beta_2-\gamma_1\beta_4\right)\ \e^{-\kappa^{-1}\int \alpha_2(z)} dz\, ,\\
Q_3&=
\begin{cases}
\displaystyle \int \e^{-\kappa^{-1}\int \alpha_3(z)} dz & \mathrm{if}\ f=f_1\, ,\\
\displaystyle \int \gamma_2\ \e^{-\kappa^{-1}\int \alpha_3(z)} dz & \mathrm{if}\ f=f_2\, ,\\
\displaystyle \int \gamma_4\ \e^{-\kappa^{-1}\int \alpha_3(z)} dz & \mathrm{if}\ f=f_3
\end{cases}
\end{aligned}
\end{align}
and
\begin{align} \nonumber
\widetilde{Q}_1&=\int \left(\beta_1+\frac{1}{2}\gamma_2\beta_4\right)\ \e^{-\kappa^{-1}\int \alpha_1(z)} dz\, ,\\
\widetilde{Q}_2&=\int \left(\beta_2-\frac{1}{2}\gamma_1\beta_4\right)\ \e^{-\kappa^{-1}\int \alpha_2(z)} dz\, ,\\
\widetilde{Q}_3&=
\begin{cases}
\displaystyle \int \e^{-\kappa^{-1}\int \alpha_3(z)} dz & \mathrm{if}\ f=f_1\, ,\\
\displaystyle \int \gamma_2\ \e^{-\kappa^{-1}\int \alpha_3(z)} dz & \mathrm{if}\ f=f_2\, ,\\
\displaystyle \int \left(\frac{1}{2}\gamma_1\gamma_2-\gamma_4\right)\ \e^{-\kappa^{-1}\int \alpha_3(z)} dz & \mathrm{if}\ f=f_3\, . \nonumber
\end{cases}
\end{align}

\bigskip

\noindent {\bf $\frac{1}{2}\Z$-graded case.} 
We have another good grading for $f_3$:
\begin{align}
\setlength{\unitlength}{1mm}
\begin{picture}(0,0)(20,10)
\put(10,10){\circle{2}}
\put(9,13){\footnotesize$\frac{1}{2}$}
\put(9,5){\footnotesize$\alpha_1$}
\put(11,10.3){\line(1,0){8}}
\put(20,10){\circle{2}}
\put(19,13){\footnotesize$0$}
\put(19,5){\footnotesize$\alpha_2$}
\put(21,10.3){\line(1,0){8}}
\put(30,10){\circle{2}}
\put(29,13){\footnotesize$\frac{1}{2}$}
\put(29,5){\footnotesize$\alpha_3$}
\put(32.5,9){.}
\end{picture}
\end{align}
If $k$ is generic,
\begin{align}
V^k(\slf(4))\simeq\Ker S_1\cap\Ker S_2\cap\Ker S_3\subset M_{\slf(4)}\otimes\pi\, ,
\end{align}
where
\begin{align} \nonumber
S_1&=\int \left(\beta_1+\gamma_1\beta_4+\gamma_5\beta_6\right)\ \e^{-\kappa^{-1}\int \alpha_1(z)} dz\, ,\\
S_2&=\int \beta_2\ \e^{-\kappa^{-1}\int \alpha_2(z)} dz,\\
S_3&=\int \left(\beta_3-\gamma_2\beta_5-\gamma_4\beta_6\right)\ \e^{-\kappa^{-1}\int \alpha_3(z)} dz\, . \nonumber
\end{align}
Applying the Drinfeld-Sokolov reduction, we have
\begin{align}
\W^k(\slf(4),\mathrm{minimal})
\simeq\Ker Q_1\cap\Ker Q_2\cap\Ker Q_3\subset \langle\beta_1,\gamma_1,\beta_2,\gamma_2,\beta_3,\gamma_3\rangle_{\mathrm{VA}}\otimes\pi\, ,
\end{align}
where
\begin{align} \label{QSCAh}
\begin{aligned}
Q_1&=\int \left(\gamma_1-\gamma_2\beta_3\right)\ \e^{-\kappa^{-1}\int \alpha_1(z)} dz\, ,\\
Q_2&=\int \beta_2\ \e^{-\kappa^{-1}\int \alpha_2(z)} dz\, ,\\
Q_3&=\int \left(\gamma_3+\gamma_2\beta_1\right)\ \e^{-\kappa^{-1}\int \alpha_3(z)} dz\, .
\end{aligned}
\end{align}

\subsection{Screening operators of $\mathcal{W}^k(\mathfrak{so}(5),f)$}
\label{sec:screeningso5}

Consider the case $\mathfrak{g}=\mathfrak{so}(5)$. Then
\begin{align}
\Pi = \{ \alpha_1, \alpha_2 \} \, ,\quad
\Delta_+ = \{ \alpha_1, \alpha_2, \alpha_1+\alpha_2, \alpha_1+2\alpha_2 \} \, .
\end{align}
Hence $M_{\mathfrak{so}(5)}$ is isomorphic to $4$ tensor products of the $\beta\gamma$-system. Set
\begin{align*}
&\beta_1=\beta_{\alpha_1}\, ,\quad
\beta_2=\beta_{\alpha_2}\, ,\quad
\beta_3=\beta_{\alpha_1+\alpha_2}\, ,\quad
\beta_4=\beta_{\alpha_1+2\alpha_2}\, ,\\
&\gamma_1=\gamma_{\alpha_1}\, ,\quad
\gamma_2=\gamma_{\alpha_2}\, ,\quad
\gamma_3=\gamma_{\alpha_1+\alpha_2}\, ,\quad
\gamma_4=\gamma_{\alpha_1+2\alpha_2}\, .
\end{align*}
Here and further on, we often drop the indeterminate $z$, e.g., we shall write $\beta$ in place of $\beta(z)$. Wakimoto free field realizations of $V^k(\mathfrak{so}(5))$ are the following embeddings
\begin{align}
V^k(\mathfrak{so}(5))\hookrightarrow M_{\mathfrak{so}(5)}\otimes\pi\, ,
\end{align}
where
\begin{align}
M_{\mathfrak{so}(5)}&=\langle\beta_i,\gamma_i\mid i=1,\ldots,4 \rangle_{\mathrm{VA}}\, ,\quad
\beta_i(z)\gamma_j(w)\sim-\frac{\delta_{i,j}}{z-w}\, ,\\
\pi&=\langle\alpha_i\mid i=1,2 \rangle_{\mathrm{VA}}\, ,\quad
\alpha_i(z)\alpha_j(w)\sim\frac{(\alpha_i|\alpha_j)}{(z-w)^2}\, .
\end{align}
If $k$ is generic, the image may be described as the common kernel of $2$ screening operators $S_1$, $S_2$. Given a nilpotent element $f$ of $\mathfrak{so}(5)$ and a good grading on $\mathfrak{so}(5)$ for $f$, these $S_i$ induce screening operators $Q_1$, $Q_2$ of $\mathcal{W}^k(\mathfrak{so}(5),f)$ via the Drinfeld-Sokolov reduction. Indeed, we have several choice of $S_i$, see below. If we choose suitable $S_i$ for each $i$ with respect to $f$, we can compute $Q_i$ explicitly.

\smallskip

Using these reduction techniques, we give examples of screening operators $Q_i$ of $\mathcal{W}^k(\mathfrak{so}(5),f)$. In $\mathfrak{so}(5)$, there exist $4$ nilpotent orbits, classified by partitions of $5$ in which even number has even multiplicity:
\begin{align}
(5)\, ,\quad
(3,1^2\, ),\quad
(2^2,1)\, ,\quad
(1^5)\, .
\end{align}
Then principal, subregular, minimal, zero nilpotent orbits correspond to $(5)$, $(3,1^2)$, $(2^2,1)$, $(1^5)$ respectively. In case $f=0$, the $\mathcal{W}^k(\mathfrak{so}(5),f)$ is just the affine vertex algebra $V^k(\mathfrak{so}(5))$. Hence we only need to consider the case $f$ belongs to one of the following nilpotent orbits: principal, subregular and minimal. We will use the following notation
\begin{align}
\kappa=\sqrt{k+3} \, .
\end{align}

\subsubsection*{Principal nilpotent}
Consider a principal nilpotent element
\begin{align}
f=f_{\alpha_1}+f_{\alpha_2}
\end{align}
with the (unique) good grading
\begin{align}
\setlength{\unitlength}{1mm}
\begin{picture}(0,0)(20,10)
\put(10,10){\circle{2}}
\put(9,13){\footnotesize$1$}
\put(9,5){\footnotesize$\alpha_1$}
\put(11,10.5){\line(1,0){8}}
\put(11,9.5){\line(1,0){8}}
\put(13.5,9){\Large$>$}
\put(20,10){\circle{2}}
\put(19,13){\footnotesize$1$}
\put(19,5){\footnotesize$\alpha_2$}
\put(22.5,9){.}
\end{picture}
\end{align}
If $k$ is generic,
\begin{align}
V^k(\mathfrak{so}(5))\simeq\Ker S_1\cap\Ker S_2\subset M_{\mathfrak{so}(5)}\otimes\pi\, ,
\end{align}
where
\begin{equation}
\begin{split}
S_1&=\int \left(\beta_1+\frac{1}{2}\gamma_2\beta_3-\frac{1}{12}\gamma_2^2\beta_4\right)\ \e^{-\kappa^{-1}\int \alpha_1(z)} dz\, ,\\
S_2&=\int \left(\beta_2-\frac{1}{2}\gamma_1\beta_3+\frac{1}{2}\left(\gamma_3+\frac{1}{6}\gamma_1\gamma_2\right)\beta_4\right)\ \e^{-\kappa^{-1}\int \alpha_2(z)} dz\, .
\end{split}
\end{equation}
Applying the Drinfeld-Sokolov reduction, we have
\begin{align}
\mathcal{W}^k(\mathfrak{so}(5))=\mathcal{W}^k(\mathfrak{so}(5),\mathrm{principal})\simeq\Ker Q_1\cap\Ker Q_2\subset \pi\, ,
\end{align}
where
\begin{equation}
\begin{split}
Q_1&=\int \e^{-\kappa^{-1}\int \alpha_1(z)} dz\, ,\\
Q_2&=\int \e^{-\kappa^{-1}\int \alpha_2(z)} dz\, .
\end{split}
\end{equation}

\subsubsection*{Subregular nilpotent}
Consider subregular nilpotent elements
\begin{align}
f_1=f_{\alpha_2},\quad
f_2=f_{\alpha_1+\alpha_2}
\end{align}
with the good grading
\begin{align}
\setlength{\unitlength}{1mm}
\begin{picture}(0,0)(20,10)
\put(10,10){\circle{2}}
\put(9,13){\footnotesize$0$}
\put(9,5){\footnotesize$\alpha_1$}
\put(11,10.5){\line(1,0){8}}
\put(11,9.5){\line(1,0){8}}
\put(13.5,9){\Large$>$}
\put(20,10){\circle{2}}
\put(19,13){\footnotesize$1$}
\put(19,5){\footnotesize$\alpha_2$}
\put(22.5,9){.}
\end{picture}
\end{align}
If $k$ is generic,
\begin{align}
V^k(\mathfrak{so}(5))\simeq\Ker S_1\cap\Ker S_2\subset M_{\mathfrak{so}(5)}\otimes\pi\, ,
\end{align}
where
\begin{equation}
\begin{split}
S_1&=\int \beta_1\ \e^{-\kappa^{-1}\int \alpha_1(z)} dz\, ,\\
S_2&=\int \left(\beta_2-\gamma_1\beta_3+\gamma_3\beta_4\right)\ \e^{-\kappa^{-1}\int \alpha_2(z)} dz\, .
\end{split}
\end{equation}
Applying the Drinfeld-Sokolov reductions associated to $f=f_1,f_2$, we have
\begin{align}
\mathcal{W}^k(\mathfrak{so}(5),\mathrm{subregular})\simeq\Ker Q_1\cap\Ker Q_2\subset \langle\beta_1,\gamma_1\rangle_{\mathrm{VA}}\otimes\pi\, ,
\end{align}
where
\begin{equation}
\begin{split}
Q_1&=\int \beta_1\ \e^{-\kappa^{-1}\int \alpha_1(z)} dz\, ,\\
Q_2&=
\begin{cases}
\displaystyle \int \e^{-\kappa^{-1}\int \alpha_2(z)} dz & \mathrm{if}\ f=f_1\, ,\\
\displaystyle \int \gamma_1\ \e^{-\kappa^{-1}\int \alpha_2(z)} dz & \mathrm{if}\ f=f_2\, .
\end{cases}
\end{split}
\end{equation}

\subsubsection*{Minimal nilpotent}
Consider a minimal nilpotent element
\begin{align}
f=f_{\alpha_1+2\alpha_2}
\end{align}
with the (unique) good grading
\begin{align}
\setlength{\unitlength}{1mm}
\begin{picture}(0,0)(20,10)
\put(10,10){\circle{2}}
\put(9,13){\footnotesize$0$}
\put(9,5){\footnotesize$\alpha_1$}
\put(11,10.5){\line(1,0){8}}
\put(11,9.5){\line(1,0){8}}
\put(13.5,9){\Large$>$}
\put(20,10){\circle{2}}
\put(19,13){\footnotesize$\frac{1}{2}$}
\put(19,5){\footnotesize$\alpha_2$}
\put(22.5,9){.}
\end{picture}
\end{align}
If $k$ is generic,
\begin{align}
V^k(\mathfrak{so}(5))\simeq\Ker S_1\cap\Ker S_2\subset M_{\mathfrak{so}(5)}\otimes\pi \, ,
\end{align}
where $S_1, S_2$ are the same as those given in the subregular nilpotent case. Applying the Drinfeld-Sokolov reductions associated to $f$, we have
\begin{align}
\mathcal{W}^k(\mathfrak{so}(5),\mathrm{minimal})\simeq\Ker Q_1\cap\Ker Q_2\subset \langle\beta_1,\gamma_1,\beta_3,\gamma_3\rangle_{\mathrm{VA}}\otimes\pi\, ,
\end{align}
where
\begin{equation}
\begin{split}
Q_1&=\int \beta_1\ \e^{-\kappa^{-1}\int \alpha_1(z)} dz\, ,\\
Q_2&=\int \left(\gamma_3-\gamma_1\beta_3 \right)\ \e^{-\kappa^{-1}\int \alpha_2(z)} dz\, .
\end{split}
\end{equation}


\section{Correlator relations for W-algebras from $\slf(N)$}
\label{sec:corrslN}

In this appendix, we derive several new correlator relations among theories with the symmetry of W-algebras associated with $\slf(N)$. 

\subsection{Reduction from affine $\slf (N)$ to QSCA}

The screening charges for the ${\slf}(N)$ WZNW model are given by
\begin{equation}\label{e1}
Q_{i+1,i}=(\beta_{i+1,i}+\sum_{j=1}^{i-1}\beta_{i+1,j}\gamma^{i,j}) e^{b\phi_i},
\end{equation}
where $i=1,2,\ldots, N-1$. We have adopted the convention in \cite{kuwahara1990conformal} and labelled all $\beta$- and $\gamma$-fields as matrix entries. For example, when $N=3$, \eqref{e1} yields two screening charges
\begin{equation}
Q_{2,1}=\beta_{2,1} e^{b\phi_1} \, , \qquad Q_{3,2}=(\beta_{3.2}+\beta_{3,1}\gamma^{2,1}) e^{b\phi_2} \, .
\end{equation}
These charges are equivalent to those used for the construction of action \eqref{actionsl3}, with $\beta_1\equiv\beta_{3,2}$, $\beta_2\equiv\beta_{2,1}$ and $\beta_3\equiv-\beta_{3,1}$. Similarly, \eqref{e1} with $N=4$ provides an equivalent version of  the screening charges of $\mathbb{Z}$-graded minimal $\slf(4)$.

In the general case of $\slf(N)$, the correlator of vertex operators constructed from the charges \eqref{e1} takes the form
\begin{align}
\corr{\prod_{\nu=1}^{M}V_\nu(z_\mu)}&=\int\mathcal{D}\Phi \,\text{exp}\Big\{-\frac 1 {2\pi}\int d^2 z\Big[\frac{G^{ij}}{2}\p\phi_{i}\pb\phi_{j}+\frac{b}{4}\sqrt{g}\mathcal{R}\sum_{i=1}^{N-1}\phi_i \nonumber \\
&\hspace{3mm} +\sum_{i=1}^{N-1}\sum_{j=1}^{i-1}\beta_ {i,j}\pb\gamma^{i,j}+\bb{i,j} \partial\gb{i,j}\Big]\Big\} \label{e4}
\times \text{exp}\Big\{\frac 1 {2\pi k}\int d^2 z\sum_{i=1}^{N-1}S_{i+1,i}\Big\}
\\&\hspace{3mm}
\times\prod_{\nu=1}^{M}\prod_{i=1}^{N-1}\prod_{j=1}^{i-1} e^{\mu_{i,j}^\nu\gamma^{i,j}-\bar{\mu}_{i,j}^\nu\gb{i,j}}\prod_{n=1}^{N-1} e^{2b(j_{n}^\nu+1)\phi_{n}(z_\nu)}\, , \nonumber 
\end{align}
where 
\begin{equation}
\mathcal{D} \Phi =\prod_{n=1}^{N-1}\mathcal{D}\phi_n \prod_{i=1}^{N-1}\prod_{j=1}^{i-1} \mathcal{D}^2\beta_{i,j}\mathcal{D}^2\gamma^{i,j}
\end{equation}
and
\begin{equation}
S_{i+1,i}=\left|\beta_{i+1,i}+\sum_{j=1}^{i-1}\beta_{i+1,j}\gamma^{i,j} \right|^2  e^{b\phi_i} \, .
\end{equation}
Moreover, $G_{ij}$ is the Cartan matrix of $\slf(N)$, and its inverse $G^{ij}$ is defined by $G_{ij} G^{jk} = \delta_{i}^{~k}$.

Integrating \wrt{} $\gamma^{N,i}$ with $i=1,2, \ldots, N-1$ yields
\begin{equation}
	\beta_{N,i}(z)=u_{N,i}\mathcal{B}_{N,i}(z, y^n_{N,i}, z_{\nu}) \quad \text{ with } \quad i=1,2, \ldots, N-1 \, , 
\end{equation}
where $\mathcal{B}_{i,j}$ is given by
\begin{equation}
	\beta_{i,j}(z)= - u _{i,j} \frac{\prod_{n=1}^{N-2}(z-y^n_{i,j})}{\prod_{\nu=1}^N(z-z_\nu)} = - u _{i,j} \mathcal{B} _{i,j}(z,y^n_{i,j},z_\nu) \, , \quad
	\sum_{\nu=1}^{N}\mu_{i,j}^{\nu}=0 \, .
\end{equation}

After the integration, the interaction term $S_{N,N-1}$ becomes
\begin{equation}\label{e3}
	S_{N,N-1}= - \left|u_{N,N-1}\mathcal{B}_{N,N-1}+\sum_{j=1}^{N-2}u_{N,j}\mathcal{B}_{N,j}\gamma^{N-1,j} \right|^2 e^{b\phi_{N-1}}\, .
\end{equation}
We remove the factor $u_{N,N-1}\mathcal{B}_{N,N-1}$ by letting
\begin{equation}
	\phi_{N-1}+\frac 1 b \ln{|u_{N,N-1}\mathcal{B}_{N,N-1}|^2}\rightarrow\phi_{N-1} \, ,
\end{equation}
upon which the interaction term \eqref{e3} becomes
\begin{equation}
	\begin{aligned}
		S_{N,N-1}&= - \left|1+u^{-1}_{N,N-1}\mathcal{B}^{-1}_{N,N-1}\sum_{j=1}^{N-2}u_{N,j}\mathcal{B}_{N,j}\gamma^{N-1,j}\right|^2 e^{b\varphi_{N-1}}\\
		&= - \left|1+\sum_{j=1}^{N-2}\gamma^{N-1,j}\right| ^2 e^{b\varphi_{N-1}} \, ,
	\end{aligned}
\end{equation}
where the following change of fields is made
\begin{equation}
	\begin{aligned}
		&u_{N,N-1}^{-1}\mathcal{B}_{N,N-1}^{-1}u_{N,j}\mathcal{B}_{N,j}\gamma^{N-1,j}\rightarrow\gamma^{N-1,j}\, ,\\
		&u_{N,N-1}\mathcal{B}_{N,N-1}u_{N,j}^{-1}\mathcal{B}_{N,j}^{-1}\beta_{N-1,j}\rightarrow\beta_{N-1,j}\, .
	\end{aligned}
\end{equation}

The interaction term $S_{N-1,N-2}$ now takes the form
\begin{equation}
	\begin{aligned}
		S_{N-1,N-2}&=\Big | u^{-1}_{N,N-1}\mathcal{B}^{-1}_{N,N-1}u_{N,N-2}\mathcal{B}_{N,N-2}\beta_{N-1,N-2}\\
		&\hspace{1cm}+u^{-1}_{N,N-1}\mathcal{B}^{-1}_{N,N-1}\sum_{j=1}^{N-3}u_{N,j}\mathcal{B}_{N,j}\beta_{N-1,j}\gamma^{N-2,j}\Big| ^2  e^{b\phi_{N-2}}\\
		&=\left|\beta_{N-1,N-2}+\sum_{j=1}^{N-3}\beta_{N-1,j}\gamma^{N-2,j}\right| ^2  e^{b\varphi_{N-2}}
	\end{aligned}
\end{equation}
where we have made the following change of fields
\begin{equation}
	\begin{aligned}
		\phi_{N-2}+\frac 1 b \ln{|u^{-1}_{N,N-1}\mathcal{B}^{-1}_{N,N-1}u_{N,N-2}\mathcal{B}_{N,N-2}|^2}&\rightarrow\varphi_{N-2}\, ,\\
		u_{N,N-2}^{-1}\mathcal{B}_{N,N-2}^{-1}u_{N,j}\mathcal{B}_{N,j}\gamma^{N-2,j}&\rightarrow\gamma^{N-2,j}\, ,\\
		u_{N,N-2}\mathcal{B}_{N,N-2}u_{N,j}^{-1}\mathcal{B}_{N,j}^{-1}\beta_{N-2,j}&\rightarrow\beta_{N-2,j}\, ,
	\end{aligned}
\end{equation}
where $j=1,2, \ldots, N-3$.

The treatment to $S_{N-2,N-3}$ and consequently all other charges follows similarly as to $S_{N-2,N-2}$, during which we have made the following change of fields
\begin{equation}
	\begin{aligned}
		\phi_i+\frac 1 b \ln{|u_{N,i}\mathcal{B}_{n,i}u^{-1}_{N,i+1}\mathcal{B}^{-1}_{N,i+1}|^2}&\rightarrow\phi_i\, ,\\
		u_{N,i}^{-1}\mathcal{B}_{N,i}^{-1}u_{N,j}\mathcal{B}_{N,j}\gamma^{i,j}&\rightarrow\gamma^{i,j}\, ,\\
		u_{N,i}\mathcal{B}_{N,i}u_{N,j}^{-1}\mathcal{B}_{N,j}^{-1}\beta_{i,j}&\rightarrow\beta_{i,j}
	\end{aligned}
\end{equation}
with $i=1,2,\ldots, N-3$ and $j=1,2,\ldots,i-1$. The correlation function then becomes
\begin{equation}
	\begin{aligned}
		&\corr{\prod_{\nu=1}^{M}V_\nu(z_\mu)}= |\Theta_M|^2 \prod_{i=1}^{N-1}\delta^{(2)}\left(\sum_{v=1}^N\mu_{N-1,i}^\nu\right)\\
		&\quad\times\corr{\prod_{n=1}^M\tilde{V_\nu}(z_\nu)\prod_{n=1}^{M-2}\prod_{i=1}^{N-2}\prod_{j=1}^{i-1}\tilde{V}^{(1)}(y_{N,i}^n)\tilde{V}^{(2)}(y_{N,i+1}^n)\tilde{V}^{(3)}(y_{N,N-1}^n)\tilde{V}^{(4)}(y_{N,j}^n)}\, ,
	\end{aligned}
\end{equation}
where the correlator on the right-hand side is computed with the action
\begin{align}
		S&=\frac 1 {2\pi}\int d^2 z\Big[\frac{G^{ij}}{2}\p\phi_{i}\pb\phi_{j}+\frac{\sqrt{g}\mathcal{R}}{4}\left(\sum_{n=1}^{N-2}b\phi_n+(b+b^{-1})\phi_{N-1} \right) +\sum_{i=2}^{N-1}\sum_{j=1}^{i-1}\beta_{i,j}\pb\gamma^{i,j}+\bar{\beta}_{i,j}\partial\bar{\gamma}^{i,j}\Big] \nonumber \\
		&\hspace{1cm}-\frac{1}{2\pi k} \int d^2 z\Big[\sum_{n=1}^{N-2}S_{i+1,i} - \tilde{S}_{N,N-1}\Big]
\end{align}
with
\begin{equation}
	\tilde{S}_{N,N-1}=\left|1+\sum_{i=1}^{N-2}\gamma^{N-1,i}\right|^2  e^{b\varphi_n}\, .
\end{equation}
The vertex operators are found to be
\begin{equation}
	\begin{aligned}
		&\tilde{V}_{\nu}(z_{\nu})=e^{\sum_{i=1}^{N-1}\sum_{j=1}^{i-1} ( \mu_{i,j} '{}^{ \nu}\gamma^{i,j}-\bar{\mu}_{i,j} '{}^{ \nu}\bar{\gamma}^{i,j})}e^{\sum_{i=1}^{N-1}2b(j_i^\nu+1)\phi_i+\phi^{N-1}/b} \, , \\
		&\tilde{V}^{(1)}(y_{N,i}^n)=e^{\phi^i/b+X_{i,j}+\bar{X}_{i,j}} \, , \quad
		\tilde{V}^{(2)}(y_{N,i+1}^n)=e^{\phi^{i+1}/b} \, , \\
		&\tilde{V}^{(3)}(y_{N,N-1}^n)=e^{-\phi^{N-1}/b+X_{N-1,j}+\bar{X}_{N-1,j}}\, ,\quad
		\tilde{V}^{(4)}(y_{N,j}^n)=e^{-X_{i,j}-\bar{X}_{i,j}-X_{N-1,j}-\bar{X}_{N-1,j}}
	\end{aligned}
\end{equation}
where in $\tilde{V}_{\nu}(z_{\nu})$
\begin{equation}
	\mu_{i,j} '{}^{ \nu} =\frac{u_{N,i}\mu_{N,i}^\nu}{u_{N,j}\mu_{N,j}^\nu}\mu_{i,j}^{\nu} \, . \end{equation}


\subsection{Reduction from affine $\slf(N)$ to W$_N$-algebra}
To obtain the correlation function of the W$_N$-algebra from that of ${\slf}(N)$, we first set $\mu^{\nu}_{i,j}=0$ with $i=3,4, \ldots, N$ and $j=1,2,\ldots,i-2$, while integrating \wrt{} $\gamma^{i,j}$. This yields $\beta_{i,j}=0$. The correlation function now takes the form
\begin{align}
		\corr{\prod_{\nu=1}^{M}V_\nu(z_\mu)}&=\int\mathcal{D} \Phi \,\text{exp}\Big\{-\frac 1 {2\pi}\int d^2 z\Big[\frac{G^{ij}}{2}\p\phi_{i}\pb\phi_{j}+\frac{b}{4}\sqrt{g}\mathcal{R}\sum_{i=1}^{N-1}\phi_i \nonumber \\
		&\hspace{0mm}+\sum_{i=1}^{N-1}\beta_ {i+1,i}\pb\gamma^{i+1,i}+\bb{i+1,i} \partial\gb{i+1,i}\Big]\Big\}\times \text{exp}\Big\{\frac 1 {2\pi k}\int d^2 z\sum_{i=1}^{N-1}\left|\beta_{i+1,i }\right|^2 e^{b\phi_i}\Big\} \nonumber \\
		&\hspace{3mm}\times\prod_{i=1}^{N-1} e^{\mu_{i+1,i}^\nu\gamma^{i+1,i}-\bar{\mu}_{i+1,i}^\nu\gb{i+1,i}}\prod_{n=1}^{N-1}e^{2b(j_{n}^\nu+1)\phi_{n}(z_\nu)}\, ,
\label{e5}
\end{align}
where
\begin{equation}
	\mathcal{D}\Phi =\prod_{n=1}^{N-1}\mathcal{D}\phi_n\prod_{i=1}^{N-1} \mathcal{D}^2\beta_{i+1,i}\mathcal{D}^2\gamma^{i+1,i} \, .
\end{equation}

Integrating over $\gamma^{i+1,i}$, where $i=1,2,\ldots,N-1$,  \eqref{e5} yields analogous relations on $\mathcal{B}_{i+1,i}$ as in the previous case. The interaction terms now are  $| u_{i+1.i}\mathcal{B}_{i+1,i} |^2  e^{b\phi_i}$, whose factor $u_{i+1.i}\mathcal{B}_{i+1,i}$ can be absorbed by a change of $\phi_i(z)$:
\begin{equation}
	\phi_{i}(z)+\frac 1 b \ln{\left|u_{i+1,i}\mathcal{B}_{i+1,i}\right|^2}\rightarrow\phi_i(z)\, .
\end{equation}
This yields final form of the W$_{N}$-correlator:
\begin{equation}\label{ap1}
	\left \langle \prod^M_{\nu = 1} V_\nu (z_\nu) \right \rangle = |\Theta_M|^2 \delta ^{(2)} \left(\sum_{\nu=1}^N\mu^\nu_{i+1,i}\right) \left \langle \prod^M_{\nu = 1} \tilde V_\nu (z_\nu) \prod_{n=1}^{M-2}\tilde{V}_b(y_{i+1,i}^n)\right \rangle \, ,
\end{equation}
where the action associated to the correlator on the right-hand side is 
\begin{equation}
	S = \frac{1}{2 \pi} \int d ^2 z \Big[ \frac{G_{ij}}{2} \partial \phi^i \bar \partial \phi^j + \frac{\sqrt{q} \mathcal{R}}{4} \left(b+\frac 1 b\right)\sum_{i=1}^{N-1}\phi_i +  \frac { 1 }{k} \sum_{i=1}^{N-1} e^{b\phi_i} \Big] \, .
\end{equation}
The vertex operators on the right-hand side of \eqref{ap1} are given by
\begin{equation}
	\tilde{V}_\nu(z_\nu)=e^{\sum_{i=1}^{N-1}2b(j_i^\nu+1)\phi_i+\phi^i/b}, \quad 
	\tilde{V}_b(y_{i+1,i}^n)=e^{-\sum_{i=1}^{N-1}\phi^i/b}.
\end{equation}
And the prefactor is found to be
\begin{equation}
	\begin{aligned}
		\Theta_M&=\prod_{i=1}^{M-1}\prod_{j<i}\prod_{\mu=1}^M\prod_{\nu<\mu}\prod_{m,n=1}^{M-2}u_{i+1,i}^2\left(\frac{(y_{i+1,i}^n-y_{j+1,j}^m)(z_\mu-z_\nu)}{(z_\nu-y_{i+1,i}^n)}\right)^{G^{ij}/b^2}\\
		&\hspace{4cm}\times\prod_{n'<n}(y^n_{i+1,i}-y^{n'}_{i+1,i})^{G^{ii}/b^2} \, .
	\end{aligned}
\end{equation}

\subsection{Reduction from W$^{(2)}_N$-algebra to W$_N$-algebra}
The subregular $\slf(N)$ is formulated by $\phi_i(z)$ and one pair of free ghosts $\beta(z)$ and $\gamma(z)$. The vertex operator correlation function takes the form
\begin{align}
		\corr{\prod_{\nu=1}^{M}V_\nu(z_\mu)}&=\int\mathcal{D}\Phi \,\text{exp}\Big\{-\frac 1 {2\pi}\int d^2 z\Big[\frac{G^{ij}}{2}\p\phi_{i}\pb\phi_{j}+\frac{1}{4}\sqrt{g}\mathcal{R}\left(b \phi^1+(b+b^{-1})\sum_{i=2}^{N-1}\phi^i\right) \nonumber \\
		&\hspace{1cm}+\beta\pb\gamma+\bb{} \partial\gb{}\Big]\Big\}\times \text{exp}\left\{\frac 1 {2\pi k}\int d^2 z\left(|\beta|^2 e^{b\phi_1} - \sum_{i=2}^{N-1} e^{b\phi_i}\right)\right\} \nonumber \\
		&\hspace{3mm}\times\prod_{\nu=1}^{M} e^{\mu^{\nu}\gamma-\bar{\mu}^{\nu}\gb{}}\prod_{i=1}^{N-1} e^{2b(j_{i}^\nu+1)\phi_{i}(z_\nu)}\, , \label{e6}
\end{align}
where
\begin{equation}
	\mathcal{D} \Phi =\prod_{i=1}^{N-1}\mathcal{D}\phi_i\mathcal{D}^2\beta\mathcal{D}^2\gamma.
\end{equation}

Following an analogous procedure as before, we integrate \eqref{e6} \wrt{} $\gamma$ and change $\phi_1$ by 
\begin{equation}
	\phi_1(z)+\frac 1 b\ln{|u\mathcal{B}|^2}\rightarrow\phi_1(z) \, .
\end{equation}
The W$_{N}$-correlator takes the form
\begin{equation}
	\left \langle \prod^M_{\nu = 1} V_\nu (z_\nu) \right \rangle = |\Theta_M|^2 \delta ^{(2)} \left(\sum_{\nu=1}^M \mu^\nu \right) \left \langle \prod^M_{\nu = 1} \tilde V_\nu (z_\nu) \prod_{n=1}^{M-2}\tilde{V}_b(y^n)\right \rangle \, ,
\end{equation}
where the action associated to the correlator on the right-hand side is 
\begin{equation}
	S = \frac{1}{2 \pi} \int d ^2 z \Big[ \frac{G_{ij}}{2} \partial \phi^i \bar \partial \phi^j + \frac{\sqrt{g} \mathcal{R}}{4} \left(b+\frac 1 b\right)\sum_{i=1}^{N-1}\phi^i + \frac { 1 }{k} \sum_{i=1}^{N-1} e^{b\phi_i} \Big] \, .
\end{equation}
The vertex operators are given by
\begin{equation}
	\tilde{V}_\nu(z_\nu)=e^{\sum_{i=1}^{N-1}2b(j_i^\nu+1)\phi_i+\phi^1/b}\, , \quad 
	\tilde{V}_b(y^n)=e^{-\phi^1/b}\, .
\end{equation}
The prefactor is
\begin{equation}
	\Theta_M=\alpha\prod_{n=1}^{M-2}\prod_{n'<n}\prod_{\mu=1}^M\prod_{\nu<\mu}\left[\frac{(y^n-y^{n'})(z_\mu-z_\nu)}{(z_\nu-y^n)}\right]^{G^{11}/b^2}\, ,
\end{equation}
where 
\begin{equation}
	\alpha=u^{2(1+\frac 1 {b^2})\sum_{i=2}^{N-1}G^{i1}+G^{11}}\, .
\end{equation}


\providecommand{\href}[2]{#2}\begingroup\raggedright\endgroup

\end{document}